\documentclass[fleqn,usenatbib]{mnras}
\usepackage[T1]{fontenc}
\usepackage{multirow}
\DeclareRobustCommand{\VAN}[3]{#2}
\let\VANthebibliography\thebibliography
\def\thebibliography{\DeclareRobustCommand{\VAN}[3]{##3}\VANthebibliography}
\usepackage{graphicx}	
\usepackage{amsmath}	
\usepackage{amssymb}
\usepackage[dvipsnames]{xcolor}
\usepackage{newtxtext,newtxmath,subfig}
\def\Fig{\mbox{Figure~}}

\def\Tab{\mbox{Table~}}

\def\Sec{\mbox{Section~}}

\def\Re{\mbox{$R_{\rm e}$}}
\def\Mauto{\mbox{{\tt MAG\_AUTO\_r}}}

\title[LIVE: Lenses In VOICE]{Lenses In VoicE (LIVE): Searching for strong gravitational lenses in the VOICE@VST survey using Convolutional Neural Networks}

\author[F.~Gentile~et~al.]
{Fabrizio~Gentile,$^{1}$
Crescenzo~Tortora,$^{2}$
Giovanni~Covone,$^{1,2,3}$\thanks{E-mail:giovanni.covone@unina.it}
Léon~V.~E.~Koopmans,$^{4}$
Chiara~Spiniello,$^{2,5}$
\newauthor
Zuhui~Fan,$^{6}$
Rui~Li,$^{7}$
Dezi~Liu,$^{6}$
Nicola~R.~Napolitano,$^{2,7}$
Mattia~Vaccari$^{8,9}$
and Liping Fu$^{10}$
\\\\
$^{1}$Dipartimento di Fisica "Ettore Pancini",
Universit\`{a} di Napoli Federico II, Compl. Univ. Monte S.
Angelo, 80126 - Napoli, Italy\\
$^{2}$INAF -- Osservatorio Astronomico di
Capodimonte, Salita Moiariello, 16, 80131 - Napoli, Italy\\
$^{3}$NFN, Sezione di Napoli, C.U. Monte S. Angelo, Via Cinthia, I-80126 Napoli, Italy\\
$^{4}$Kapteyn Astronomical Institute, University of Groningen, P.O Box 800, 9700 AV Groningen, The Netherlands \\
$^{5}$Sub-Dep. of Astrophysics, Dep. of Physics, University of Oxford, Denys Wilkinson Building, Keble Road, Oxford OX1 3RH, UK \\
$^{6}$South-Western Institute for Astronomy Research, Yunnan University, Kunming 650500, P.R.China \\
$^{7}$School of Physics and Astronomy, Sun Yat-sen University Zhuhai Campus, Daxue Road 2, 519082—Tangjia, Zhuhai, Guangdong, China \\
$^{8}$Inter-university Institute for Data Intensive Astronomy,Department of Physics and Astronomy, University of the Western Cape,\\ Robert Sobukwe Road,
7535 Bellville, Cape Town, South Africa \\
$^{9}$INAF - Istituto di Radioastronomia, via Gobetti 101, 40129 Bologna, Italy\\
$^{10}$The Shanghai Key Lab for Astrophysics, Shanghai Normal University, 100 Guilin Road, Shanghai 200234, P.R.China \\
}

\date{Accepted XXX. Received YYY; in original form ZZZ}

\pubyear{2021}

\begin{document}
\label{firstpage}
\pagerange{\pageref{firstpage}--\pageref{lastpage}}
\maketitle

\begin{abstract}
We present a sample of 16 likely strong gravitational lenses identified in the \textit{VST Optical Imaging of the CDFS and ES1 fields} (VOICE survey) using Convolutional Neural Networks (CNNs). We train two different CNNs on composite images produced by superimposing simulated gravitational arcs on real Luminous Red Galaxies observed in VOICE. Specifically, the first CNN is trained on single-band images and more easily identifies systems with large Einstein radii, while the second one, trained on composite RGB images, is more accurate in retrieving systems with smaller Einstein radii. We apply both networks to real data from the VOICE survey, taking advantage of the high limiting magnitude (26.1 in the \textit{r}-band) and low PSF FWHM (0.8" in the \textit{r}-band) of this deep survey. We analyse $\sim21,200$ images with $mag_r<21.5$, identifying 257 lens candidates. To retrieve a high-confidence sample and to assess the accuracy of our technique, nine of the authors perform a visual inspection. Roughly 75\% of the systems are classified as likely lenses by at least one of the authors. Finally, we assemble the LIVE sample (Lenses In VoicE) composed by the 16 systems passing the chosen grading threshold. Three of these candidates show likely lensing features when observed by the Hubble Space Telescope. This work represents a further confirmation of the ability of CNNs to inspect large samples of galaxies searching for gravitational lenses. These algorithms will be crucial to exploit the full scientific potential of forthcoming surveys with the \textit{Euclid} satellite and the \textit{Vera Rubin Observatory}.
\end{abstract}

\begin{keywords}
gravitational lensing: strong -- galaxies: elliptical and lenticular, cD
\end{keywords}



\section{Introduction}

Gravitational lenses are astrophysical systems created when the space-time warp around foreground astrophysical objects (the lenses) deflects light rays from distant background sources. In the presence of a massive object (e.g., a galaxy or a galaxy cluster), one defines \textit{strong gravitational lensing} which produce multiple images of a distant source (when the source is a point-like object) or gravitational arcs (when the background is an extended object, such as high-\textit{z} galaxy), as predicted by \cite{Zwicky_Lensing}. 
The main observables of strong lensing (i.e., position and shape of the lensed images) strongly rely on two factors: the angular diameter distances involving observer, lens and source, and the mass distribution (baryons plus dark matter) of the lens \citep{Schneider_Lensing, bartelmann_theory}.
For these reasons, strong gravitational lensing is suitable for a wide range of astrophysical and cosmological studies, among which the estimation of the Hubble constant \citep[see e.g.][]{refsdal_hubble,Wong_Holicow2} and the measure of the dark matter fraction in early-type galaxies \citep[e.g.][]{Covone2009, Tortora_DM, treu_DM,Auger_DM,spiniello_DM}. Strong lensing has also been used to constrain the Initial Mass Function in early-type galaxies \citep[e.g.][]{Treu_IMF,Auger_IMF,barnabe_IMF,Sonnenfeld_IMF}, to identify dark matter substructures \citep[e.g.,][]{Vegetti_DM,Koopmans_DM,mao_substructure,dalal_substructure}, and to constrain cosmological models \citep[e.g.][]{Cao_cosmology,Chae_Statistics}. For a more detailed review of strong lensing applications, please refer to \citet{Treu_Lensing} and \citet{Blandford_Review}. All these analyses, however, require large samples of observed and modelled strong lenses. Unfortunately, due to the limited cross-section, strong lensing is a rare phenomenon \citep{Schneider_Lensing}. Traditionally, visual inspection used to be the main approach to lens finding \citep[see e.g.][]{Sygnet_Inspection,LeFevre_inspection}, often preceded by a spectroscopic or photometric selection of the most promising candidates \citep[e.g.][]{Bolton_SLACSFirst,Browne_CLASS,Faure_Cosmos}. However, next-generation surveys with forthcoming facilities such as the \textit{Euclid} satellite \citep{Laureijs_Euclid}, the \textit{Vera Rubin Observatory} \citep{LSST} and the \textit{Chinese Space Station} \citep{Gong_CSSOC} are expected to retrieve $\sim10^5$ strong lenses in $\sim10^9$ observed galaxies \citep{Collett_LensPop}. The high number of strong lenses identified in these surveys will allow new statistical studies about the strong lensing phenomenon. \citep[see e.g.][]{Sonnenfeld_Statistical,Oguri_statistical}. A complete review of the possible applications of large samples of strong lenses can be found in the White Papers provided by the LSST collaboration \citep{LSST} and the Euclid Collaboration (Bergamini et al., in prep.)

It is then clear that we need more efficient methods to analyse the large amounts of data produced by these facilities, reducing the need for visual inspection (a time-consuming procedure and prone to multiple biases). In the last years, several alternative methods have been developed. These spanned from crowd science \citep[e.g.][]{Marshall_Crowd} to automated source extraction \citep[e.g.][]{More_CFHTLS}. Among these, machine learning-based algorithms appeared to be the most efficient and reliable (see, for example, the results of the first \textit{strong lens finding challenge}, \citealt{Metcalf_Challenge}). Convolutional Neural Networks (CNNs, \citealt{Lecun_ANN,LeCun_CNNReview}) represent a special class of these algorithms. These networks are designed to resemble animal and human visual cortex and are currently the state-of-the-art in image recognition and classification \citep[see e.g.][]{Russakovsky_Competition}. CNN-based lens-finders have already been employed to search for galaxy-galaxy strong lenses in several wide sky surveys such as the \textit{Kilo-Degree Survey} (KiDS, \citealt{Petrillo_1,Petrillo_2,Petrillo_3,Li_KiDS}), the \textit{Dark Energy Survey }(DES, \citealt{Jacobs_DES,Jacobs_DES2}) the \textit{Pan-STARRS survey} \citep{Canameras_HoliSmokes}, the \textit{DESI survey} \citep{Huang_Desi}, and the \textit{Canada-France-Hawaii Telescope Legacy Survey} (CFHTLS, \citealt{Jacobs_CFHTLS}).

While a large amount of work has been done in analysing wide and shallow surveys, little interest has been devoted to smaller and deeper surveys. Surveys with longer exposure times and fainter limiting magnitudes are expected to retrieve more easily lenses with faint lensing features, increasing the number of identified strong lenses per square degree \citep{Collett_LensPop}. The samples of systems retrieved in these deep surveys will have higher mean redshifts (for both lenses and lensed sources). This will allow several applications to be extended to higher redshifts \citep[see e.g.,][]{treu_DM,Treu_IMF,Koopmans_DM,Vegetti_DM}. Both the \textit{Euclid} satellite and the \textit{Vera Rubin Observatory} will have, in fact, a deep survey besides their wide surveys \citep{Laureijs_Euclid,LSST}. Testing machine learning techniques on data from deep surveys is therefore crucial to exploit the full scientific potential of these forthcoming facilities.

In this paper we employ the two Convolutional Neural Networks developed in \citet{Petrillo_1,Petrillo_2} to search for strong gravitational lenses in the \textit{VST Optical Imaging of the CDFS and ES1 fields} (VOICE survey, \citealt{Vaccari_VOICE}). Both networks were already successfully employed to search for gravitational lenses in the KiDS survey \citep{Petrillo_1,Petrillo_2,Petrillo_3} and in the \textit{Fornax Deep Survey} \citep[see the preliminary results in][]{Cantiello_FDS}. Applying these CNNs to a smaller but deeper survey than KiDS, as VOICE, which has a \textit{r}-band limiting magnitude at 5$\sigma$ for point like sources of 26.1 (i.e., one magnitude deeper than KiDS; \citealt{Kuijken_DR4}),
we expect to identify a larger number of lenses per square degree than in the KiDS survey \citep{Petrillo_3,Li_KiDS,He_KiDS}. Furthermore, since the limiting magnitude of VOICE in the \textit{r}-band is comparable with the one expected for the \textit{Euclid} deep survey \citep[$\sim$26.4 at 10$\sigma$ for extended sources in the VIS band;][]{Laureijs_Euclid}, our results will be useful to predict the performances of machine-learning algorithms like CNNs on these future observations

This paper is organised as follow. In \Sec\ref{sec:Data} we briefly introduce the VOICE survey and describe the data employed to train the CNNs and to search for strong lenses. In \Sec\ref{sec:Methods} we describe the two lens finding algorithms and the procedure followed to create the training set. In \Sec\ref{sec:test} we summarise the performances of the CNNs computed applying the networks to a validation set and, in \Sec\ref{Sec:Results}, the application of the algorithms to real data from the VOICE survey. In \Sec\ref{Sec:Discussion} we present and analyse the LIVE sample (\textit{Lenses In VoicE}), comparing its size and properties with the expected number of lenses estimated with \textsc{LensPop} \citep{Collett_LensPop} and with the results found by \citet{Petrillo_3}. Finally, we summarise our conclusions in \Sec\ref{sec:conclusion}.

\begin{table}
\centering
\caption{Total exposure times (in hours) of the four VOICE-CDFS fields in the three photometric bands \textit{gri} selecting only the best exposures with PSF FWHM<1.1''. The mean seeing and the mean limiting magnitude at 5$\sigma$ for point like sources are reported in the last rows for each band (\Sec \ref{sec:Data})}
\begin{tabular}{cccc}
\hline 
 & \textit{g} & \textit{r} & \textit{i} \\ 
\hline 
CDFS-1 & 2.4 & 12.0 & 6.3 \\ 
CDFS-2 & 2.8 & 11.3 & 3.7 \\ 
CDFS-3 & 2.3 & 14.2 & 6.0 \\ 
CDFS-4 & 2.4 & 12.5 & 6.1 \\ 
\hline 
Mean Seeing & 0.6'' & 0.8'' & 0.8'' \\
Limiting Magnitude & 25.4 & 26.1 & 25.2 \\
\hline
\end{tabular} 
\label{tab:exp}
\end{table}

\section{Data from the VOICE survey}
\label{sec:Data}

The \textit{VST Optical Imaging of the CDFS and ES1 Fields} (VOICE survey, PIs: Giovanni Covone and Mattia Vaccari; \citealt{Vaccari_VOICE})
is a deep optical survey performed with the \textit{VLT survey telescope} (VST) during the INAF Guaranteed Time of Observation. The VST \citep{Capaccioli_VST} is a 2.6m optical telescope located at the ESO Paranal Observatory (Chile). Its main scientific instrument is a wide-field imager called OmegaCAM \citep{Kuijken_OmegaCAM}, which consists of a 32 CCDs grid, each made up of 4k $\times$ 2k pixels, with a field of view of about 1 deg$^2$ and a pixel size of 0.214 arcsec/pixel. The VOICE survey, once completed, will observe in the four photometric bands \textit{ugri} a sky area of $\sim$8 deg$^2$ evenly split between the \textit{Chandra Deep Field South}  \citep[CDFS;][]{Giacconi_CDFS,Tozzi_CDFS} and the \textit{European ISO Field 1} \citep[ES1;][]{Rowan_ES1,Oliver_ES1}. Several facilities already observed these regions, collecting data in different wavelengths from radio to x-rays, providing a unique set of ancillary data for these two fields \citep{Vaccari_VOICE}. This paper focuses on a 4.9 deg$^2$ area in the CDFS (RA:$3^h32^m32^s$, DEC:$-27^\cdot 48'30''$) whose VST observations took place between 2011 and 2015 and are now concluded.

The survey observing strategy consists of splitting each field in four tiles of about 1 deg$^2$. Each tile is observed several times (more than 100 exposures were taken for the \textit{r-}band observations, $\sim50$ for the other bands), reserving best observing conditions (lower seeing and darker moon phases) for the \textit{r-}band imaging. Single exposure times are 360s for the \textit{r-} and \textit{g-}bands, and 400s for the \textit{i-}band, respectively. Since observations covered about four years, image quality is not constant throughout the exposures. The PSF FWHM spans from 0.4" to 1.5" with a median value of 0.85". Images analysed in this work are obtained stacking selected exposures with PSF FWHM$<1.1$". The averaged PSF FWHMs in the final images are 0.8" for the \textit{r-} and \textit{i-}band, and 0.6" for \textit{g-}band, respectively. The total exposure time of the coadds in the \textit{r-}band spans from $11.3h$ to $14.2h$ (\Tab\ref{tab:exp}). Such long exposure times allowed us to reach a 5$\sigma$ limiting magnitude for point like sources of 26.1 in the \textit{r-}band, 25.4 in the \textit{g-}band and 25.2 in the \textit{i-}band. These deep observations were used for weak-lensing studies \citep{Fu_WL,Liu_WL}, while the multi-epoch imaging of the CDFS allowed variability selection of supernovae \citep{Cappellaro_SNe} and AGN \citep{DeCicco_AGN,Poulain_AGN}.

In this work we use VOICE data to search for strong gravitational lenses in the CDFS. There are two main reasons why these data are particularly suitable for this research. Firstly, the faint limiting magnitude makes it easier to identify strong lensing features (which are generally faint). Secondly, the low value of the PSF FWHM makes it possible to resolve lenses with small values of the Einstein Radius (i.e., the typical angular separation between arcs and deflectors) of the order of the arcsecond. This kind of lenses is generally harder to identify, but is also the most common \citep{Collett_LensPop}. Furthermore, the CDFS will be covered by the forthcoming LSST deep survey \citep{LSST}, providing a multi-epoch high-resolution follow up for our lens candidates. Finally, VOICE data are similar to the data from the \textit{Kilo-Degree Survey} (KiDS, \citealt{Kuijken_KiDS,Kuijken_DR4}), which have been already analysed by the Convolutional Neural Networks employed in this work \citep{Petrillo_1,Petrillo_2,Petrillo_3}. This will allow an interesting comparison (\Sec\ref{sec:kids}).

\begin{figure}
	\includegraphics[width=\columnwidth]{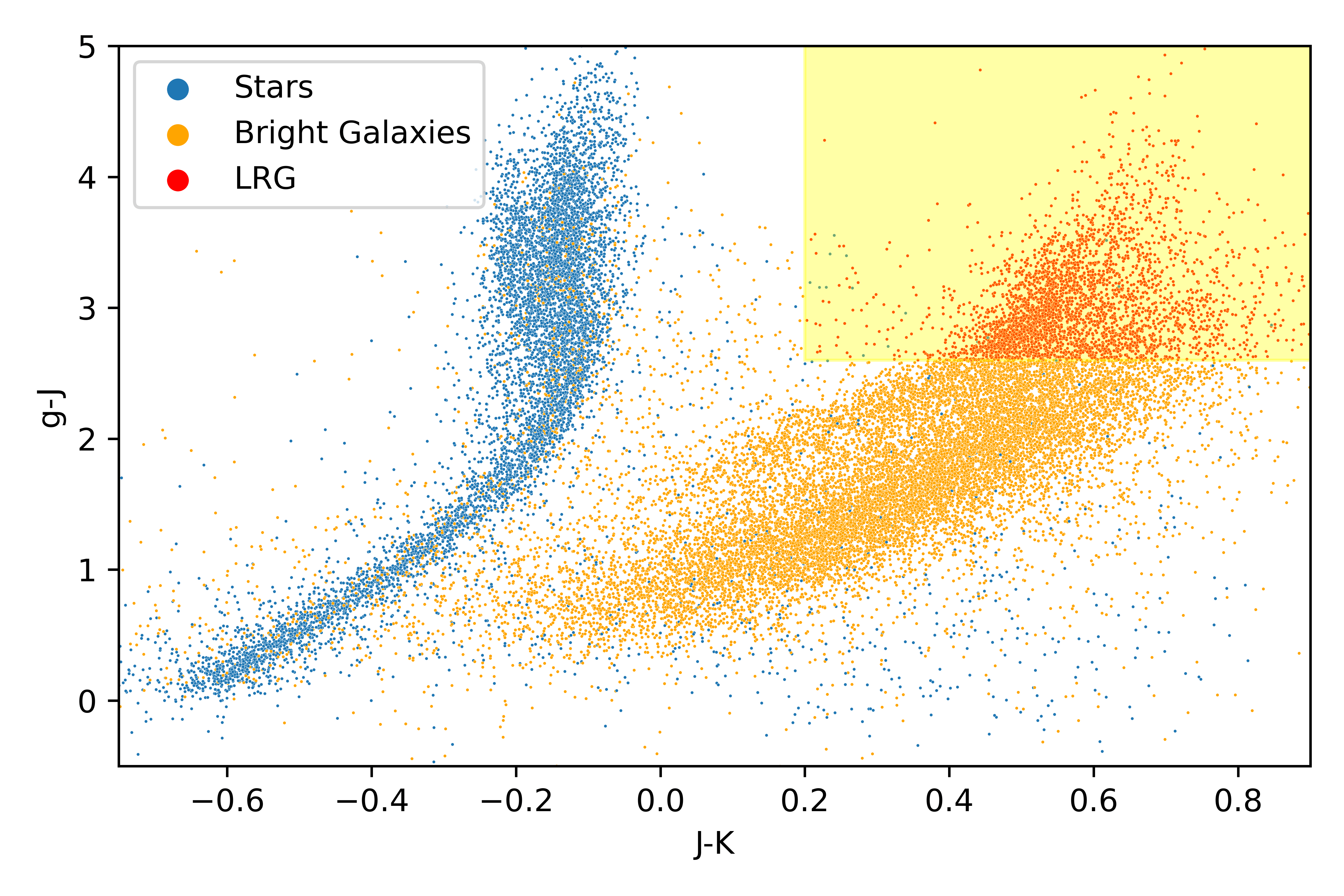}
    \caption{The $g-J$ vs $J-K$ diagram for the objects with correct photometry and \textsc{sExtractor}'s \Mauto{}<21.5. The star/galaxy separation is performed trough the \textsc{2DPhot} SG index. In the highlighted area there are the selected Luminous Red Galaxies. To generate this plot we employ NIR photometry from the VIDEO survey \citep{Jarvis_video} (\Sec \ref{sec:Sample})}
    \label{fig:LRG}
\end{figure}

\subsection{Sample Selection}
\label{sec:Sample}
The full VOICE catalogue contains 736,518 detected sources. Many of these are stars and low-mass galaxies which have a negligible strong lensing cross-section \citep{Schneider_Lensing}. Furthermore, spirals represent a small portion of the strong lenses population ($\sim$ 20\%, \citealt{Oguri_Statistics,Moller_Statistics}) and, due to their morphology, are harder to identify since the spiral arms can easily be mistaken for strong lensing features. Supplying these images to the CNN lens-finders could produce a highly contaminated candidate sample and increase network confusion during the training. We thus select galaxies with a higher probability of being strong lenses. Before operating any cut in magnitude or colour, we cross-match the VOICE catalogue with ancillary data from the VIDEO survey \citep{Jarvis_video}, obtaining photometry in the two NIR photometric bands \textit{J} and \textit{K}. We then remove from the VOICE catalogue all the objects with corrupted photometry in the photometric bands \textit{griJK} and exclude stars employing the \textsc{2DPhot} SG index \citep{LaBarbera_2dphot}. This index is particularly efficient in performing the star-galaxy separation, as can be seen in \Fig\ref{fig:LRG}. The full catalogue is thus reduced to 172,316 objects. Then, we assemble two subsets as follow:
\begin{itemize}
    \item \textit{Bright Galaxies Sample}: Lensing cross-section increases with the square of the mass of a galaxy, and therefore with luminosity \citep{Schneider_Lensing}. To select galaxies likely to be strong lenses, we select all the objects with Kron-like magnitude \Mauto{} provided by \textsc{sExtractor} \citep{Bertin_sExtractor} brighter than 21.5. The final BG sample consists of $21,216$ galaxies.
    
    \item \textit{Luminous Red Galaxies Sample}: LRGs are thought to represent most of the strong lens population \citep{Eisenstein_LRG,Oguri_LRG}. These galaxies are generally selected using the criteria of \citet{Eisenstein_LRG}. However, such selection would limit our sample to too few objects to successfully train the CNN. We thus employ a slightly modified version of the colour cut presented in \citet{Tortora_UCMG}. Starting from the BG sample ($r<21.5$), we select galaxies in the colour range:
    \begin{equation*}
    \centering
        \begin{cases} g-J>2.6\\
                J-K>0.2,\\
        \end{cases}
    \end{equation*}
shown in \Fig\ref{fig:LRG}.
We choose this \textit{g-J} threshold through a visual inspection of the sample to qualitatively assess the fraction of blue and star-forming galaxies. The chosen threshold results to be higher than the one adopted in \citet{Tortora_UCMG}. The final LRG sample consists of  $3,450$ galaxies.
\end{itemize}

\section{Methods}
\label{sec:Methods}
In this section we briefly introduce the Convolutional Neural Networks employed to search for strong gravitational lenses in the VOICE survey. We describe the procedure followed to create the training set, to simulate mock gravitational lenses, and to train the CNNs.

\subsection{Convolutional Neural Networks}
\label{sec:CNN}
Artificial Neural Networks (ANNs; \citealt{Lecun_ANN}) are among the most popular supervised machine learning algorithms. Their architecture reflects the natural neural networks, centre of animal (and human) learning process. ANNs look for the highly complex relationship between input data (e.g. galaxy images) and the target value (e.g. the probability of being a strong gravitational lens). According to the Universal Approximation Theorem \citep{Hornik_Theorem}, ANNs try to approximate this relationship applying several non-linear functions to the input data. In classic ANNs, the input data pass through different layers. Each layer is made up of multiple neurons, each of which takes as input a vector $x_i$ from the previous layer and returns a scalar $y$ given by
\begin{equation}
\centering
y = f \left( \sum _{i=0}^{N} x_i \cdot w_i + b \right) 
\end{equation}
The non-linear function $f$ is called \textit{activation function}, $w_i$ are free parameters called \textit{weights} and $b$ is the \textit{bias}. During the training phase, an ANN inspects several labelled (i.e. pre-classified) examples and \textit{learns} the classification scheme. Learning is achieved gradually adjusting the weights $w_i$ to minimise the difference between predicted and actual target value \citep{Rumelhart_Learning}. This difference is measured, for example, through a \textit{loss function} such as binary cross-entropy \citep[e.g.][]{Goodfellow_ML}:
\begin{equation}
\label{eq:cross_entropy}
    H=-t \log(p)-(1-t)\log(1-p)
\end{equation}
where \textit{t} is the target value and \textit{p} is the predicted one. Convolutional Neural Networks (CNNs \citealt{LeCun_CNNReview}) are a noteworthy subclass of ANNs. These algorithms use convolutional kernels to extract features, maintaining the 2D topology of input data. Thanks to this property, CNNs are usually employed in images classification problems where they can achieve even higher accuracy than humans (\citealt{Russakovsky_Competition, He_Human}, see \citealt{Metcalf_Challenge} and \citealt{Becker_CNNs} for some astrophysical examples). 

In this work, we use the two CNNs developed in \citet{Petrillo_1,Petrillo_2} that implement a \textit{ResNet}-like architecture\footnote{https://github.com/CEnricoP/cnn\_strong\_lensing} with four residual blocks of two convolutional layers \citep{He_ResNet}. Further details on the architecture of the CNNs can be found in \citet{Petrillo_2}. The first CNN (\textit{single-band CNN} hereafter) takes as input only \textit{r-}band images. We choose this photometric band because of its better image quality (see \Sec\ref{sec:Data}) which simplifies identifying strong lensing morphological features. The second CNN (\textit{three-band CNN}) takes as input composite images obtained combining \textit{gri} data trough the \textsc{HumVi} opensource library\footnote{https://github.com/drphilmarshall/HumVI} \citep{Marshall_HumVi}. The \textit{three-band CNN}, analysing RGB images, can recognise gravitational lenses trough the colour gradient between the redder deflecting galaxy and the bluer alleged gravitational arc. 

Both CNNs take as input 101$\times$101 pixels$^2$ stamps (equivalent to 20$\times$20 arcsec$^2$) and give as output a single \textit{p-value} in the range $[0,1]$, related to the probability that the object in the image is a strong gravitational lens \citep{Saerens_pvalue}. As already done in \citet{Petrillo_1}, we choose the size of the stamp to be small enough to speed up training phase, to exclude environment galaxies that could confuse the network, but large enough to include the largest Einstein radius expected for galaxy-galaxy lensing \citep{Collett_LensPop}. The CNNs are implemented in \textsc{Python} 3.7 using the opensource libraries \textsc{Keras}\footnote{https://keras.io/} \citep{Chollet_keras} and \textsc{TensorFlow}\footnote{https://www.tensorflow.org/} \citep{Abadi_TensorFlow}. Both networks minimise the binary cross-entropy (Eq.~\ref{eq:cross_entropy}) using the \textsc{Adam} optimiser \citep{Kingma_Adam}.

\subsection{Creating the Training Set}
From a machine learning perspective, identifying strong gravitational lenses is a two-classes classification problem. We can successfully address such issue using ANNs through appropriate training. Training these algorithms requires feeding examples from the two classes (i.e. lenses and non-lenses) to the ANNs. To successfully train our CNNs (each having $\sim 10^7$ free parameters to estimate) we need a vast pre-classified training set. However, strong gravitational lensing is a rare phenomenon. Currently, the \textit{Sloan Lens ACS Survey} (SLACS; \citealt{Bolton_SLACSFirst}) provides the largest catalogue of confirmed strong lenses comprising just 118 objects \citep{Shu_SLACSLast}. Larger databases (e.g. the MasterLens project \footnote{masterlens.astro.utah.edu/}) reach up to $\sim 700$ lenses, but many of them still require high-resolution follow-up or spectroscopic confirmation. Furthermore, all these samples do not cover homogeneously the lensing parameter space, resulting thus unsuitable for training a CNN-based lens finder to detect all possible strong lensing configurations. With a few exceptions \citep[e.g.][]{Huang_Desi}, training this kind of classifiers requires strong lensing simulations.

\begin{table}
\centering
\caption{Range of parameters used to simulate mock gravitational arcs according to \citet{Petrillo_2}. We perform uniform sampling for all parameters except for Einstein radius and source effective radius which are sampled logarithmically  (\Sec \ref{sec:simulations})}
\begin{tabular}{ccc}
\hline 
Parameter & Range & Units  \\
\hline 
 \multicolumn{3}{c}{Lens (SIE)}  \\
\hline 
Einstein radius & 1.0 - 5.0 (log) & arcsec  \\
Axis ratio & 0-3 - 1.0 & -  \\
Major-axis angle & 0 - 180 & degrees  \\
External shear & 0 - 0.05 & -  \\
External shear angle & 0 - 180 & degrees \\ 
\hline
 \multicolumn{3}{c}{Source (Sérsic)} \\
\hline 
Effective Radius \Re\ & 0.2 - 0.6 (log) & arcsec  \\
Axis ratio & 0-3 - 1.0 & -  \\
Major-axis angle & 0 - 180 & degree  \\
Sérsic Index & 0.5 - 5.0 & -  \\
\hline 
\multicolumn{3}{c}{Sérsic Blobs (1 up to 5)}  \\
\hline 
Effective radius & $(1\% - 10\%) \Re$ & arcsec  \\
Axis ratio & 1.0 & -  \\
Major-axis angle & 0 & degrees  \\
Sérsic Index & 0.5 - 5.0 & -  \\
\hline 
\end{tabular} 
\label{tab:param}
\end{table}

\begin{figure*}
	\includegraphics[scale=0.75]{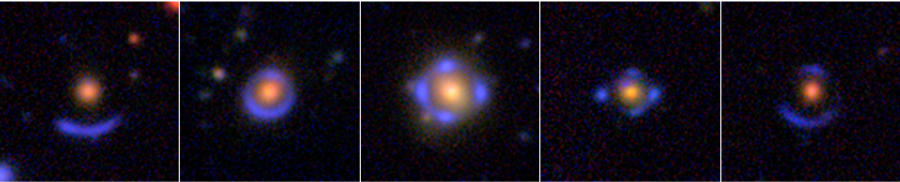}
    \caption{Examples of mock strong gravitational lenses simulated to train the \textit{three-band CNN}. All images are created superimposing a simulated gravitational arc on a real LRG observed in the VOICE survey. All images have a 20 arcsec side. Further details in \Sec\ref{sec:simulations}}
    \label{fig:Mocks}
\end{figure*}

\subsubsection{Simulating Strong Lenses}
\label{sec:simulations}
To simulate strong gravitational lenses, we can follow two different strategies: we can simulate both deflectors and gravitational arcs \citep[e.g.][]{Metcalf_Challenge,Pourrahmani_LensFlow} or we can simulate the arcs and superimpose them on real galaxy images \citep[e.g.][]{Petrillo_1,Li_KiDS}. In this work, we follow the second strategy. By doing so, we obtain realistic images (\Fig\ref{fig:Mocks}) without having to simulate sky and instrument noise nor the nearby environment or line-of-sight structures around the lens galaxy. We produce simulations using the software described in \citet{Chatterjee_PhD}. We model the mass distribution of the lens galaxies (deflectors) using a Singular Isothermal Ellipsoid model \citep[SIE,][]{Kormann_SIE,Gavazzi_SIE} with external shear \citep{Keeton_Sie}. The parameters of the model are sampled in the range used in \citet{Petrillo_2,Petrillo_3} and summarised in \Tab\ref{tab:param}.

We choose a uniform sampling for the axis ratio, inclination, shear strength, and shear angle, while we employ logarithmic sampling for the Einstein radius. By doing so, we train our CNN to identify more easily systems with small Einstein radii that are generally harder to detect but also more common \citep{Collett_LensPop}. We also simulate background lensed galaxies: we use a Sérsic brightness profile \citep{Sersic_Profile} with parameters sampled from the range in \Tab\ref{tab:param}. Similarly, we choose uniform sampling for the axis ratio, inclination, and Sèrsic index, while we employ logarithmic sampling for effective radius. To add additional structures to the matter distribution, as in \citet{Petrillo_2}, we add a Gaussian Random Field in the lens plane \citep{Ezaveh_GRF} and from one to five Sérsic components in the brightness distribution of the lensed source, to crudely mimic star-forming regions \citep{Chatterjee_Substructure}. These perturbations were shown to increase the accuracy of CNN-based lensfinders \citep{Petrillo_2,Petrillo_3} Further details on our simulation strategy can be found in \citet{Petrillo_2} and \citet{Chatterjee_PhD}.

\subsubsection{Positive training set}
\label{sec:positive}
For the \textit{single-band CNN}, we produce mock strong lenses following a slightly modified version of \citet{Petrillo_2} strategy:
\begin{enumerate}
\item We randomly select deflectors from the LRG sample (see \Sec\ref{sec:Sample});
\item We simulate 101$\times$101 pixels$^2$ stamps of gravitational arcs with the same pixel scale as the VST. We convolve them with an averaged PSF obtained by applying the PSFEx software \citep{Bertin_PSFEx} to the \textit{r}-band VOICE tiles. Differently from \citet{Petrillo_2}, we directly simulate gravitational arcs during the training phase to increase the number of strong lensing configuration examined by the CNN;
\item We normalise gravitational arcs to the deflector maximum brightness multiplied by an $\alpha$-factor in the range $[0.02,0.3]$. This factor accounts for the expected luminosity gradient between deflector and arc;
\item We coadd the two images, applying a square root stretching to enhance lensing features;
\item Finally, we normalise all pixel values to the maximum brightness in the image.
\end{enumerate}

We create images to supply to the \textit{three-band CNN} through the same procedure, with a few differences:
\begin{enumerate}
\item We simulate three copies of each arc, one for each photometric band. We convolve each arc with the corresponding averaged PSF;
\item We “colour” gravitational arcs using synthetic photometries of Late-Type Galaxies (LTG) from the COSMOS templates in the \textsc{LePhare} library \citep{Arnouts_LePhare}. These are synthetic models used to estimate photometric redshifts of galaxies in the COSMOS fields \citep{Ilbert_LePhare}. The full library contains 31 templates in total, for elliptical/S0 galaxies (8 models), spirals (11 models) and star-bursting galaxies (12 models). We select photometries of LTG and star-bursting galaxies (template index > 19) and redshift them to different values of $z$ up to $z=3$. We employ later-type templates than \citet{Petrillo_2} to increase the colour gradient between deflectors (i.e. LRGs) and lensed sources. This choice is shown to decrease the number of environment galaxies erroneously classified as arcs. 
\item To homogeneously sample colour space and to account for possible errors in the photometry, we add a random term in the range $[-0.1,0.1]$ to \textsc{LePhare} magnitudes. We also add a color-excess term $A_x=R_x E(B-V)$ to account for extinction. In this relation, $x$ is the SDSS filter considered and $R_x$ factors are taken from \citet{Yuan_Extinction}
\item We combine the three images using the \textsc{HumVi} opensource code \citep{Marshall_HumVi} which applies the Lupton’s algorithm \citep{Lupton_RGB}, performing a $sinh$ stretching instead of a more standard square-root one.
\end{enumerate}

\subsubsection{Negative training set}
\label{sec:negative}
A good lens-finder is required to produce a pure candidate sample. We thus need to teach the CNN how to recognise and exclude contaminants. Several studies \citep[e.g.][]{Petrillo_1,Petrillo_2,Petrillo_3,Li_KiDS} reported how some objects (e.g. spirals, merging galaxies, polar rings) can easily confuse CNN-based classifiers because of their morphology and colour gradient. To limit such effects, we populate our negative training set ($\sim$ 40\% of the full set) with the bluest sources in the BG sample with $g-J<2.6$ (i.e. the ones with a higher probability of being spirals or star-forming galaxies). We populate the remaining 60\% with other random galaxies in the BG sample (30\%) and LRGs from the homonymous sample (30\%). As highlighted by \citet{Petrillo_1,Petrillo_2}, we cannot exclude that a few real lenses are present in our negative sample, but their expected low number (less then 1 in a 1000) should not strongly affect our training.

\begin{figure*}
\subfloat[][]
{\includegraphics[scale=0.55]{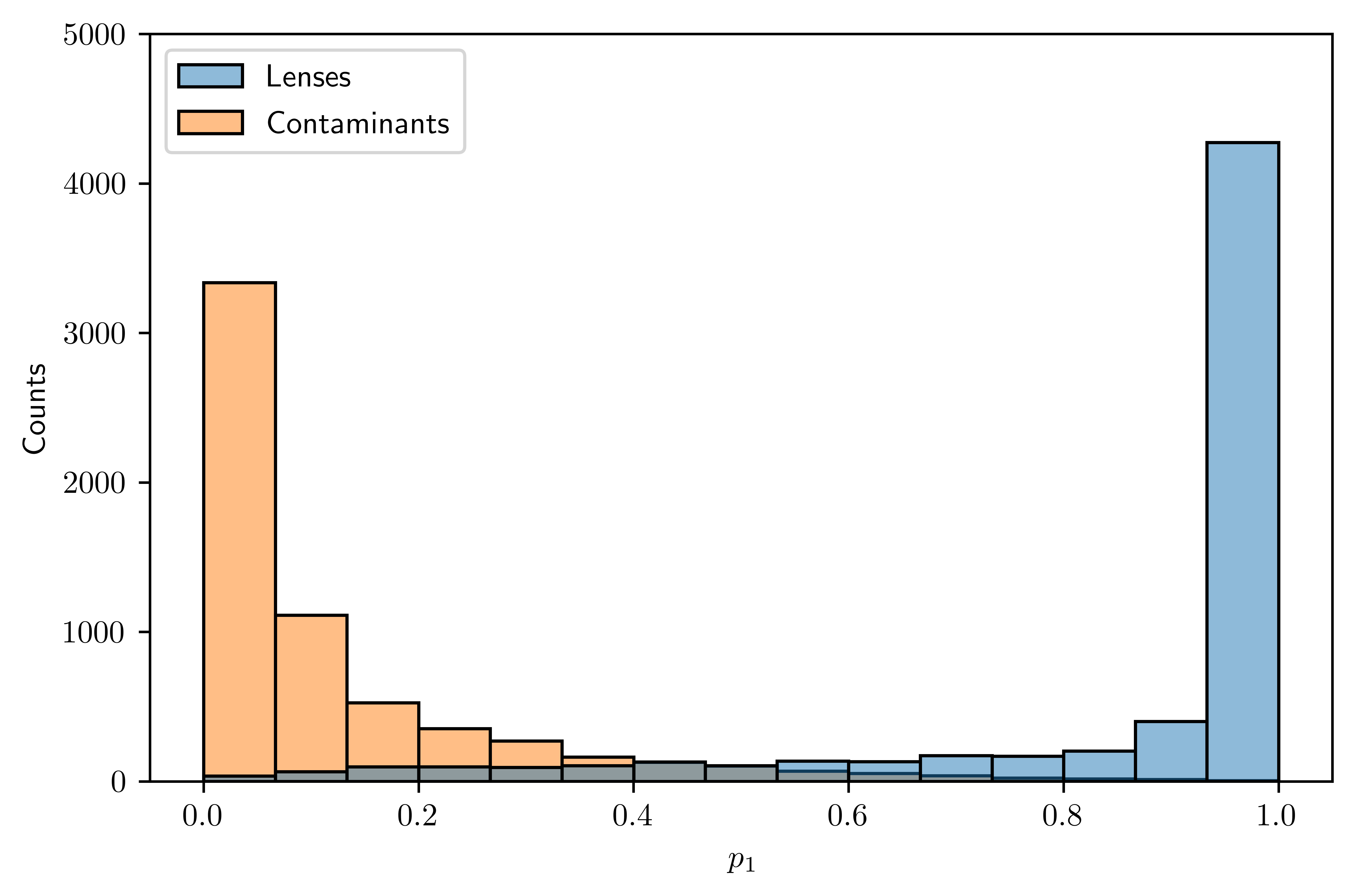}}\qquad
\subfloat[][]
{\includegraphics[scale=0.55]{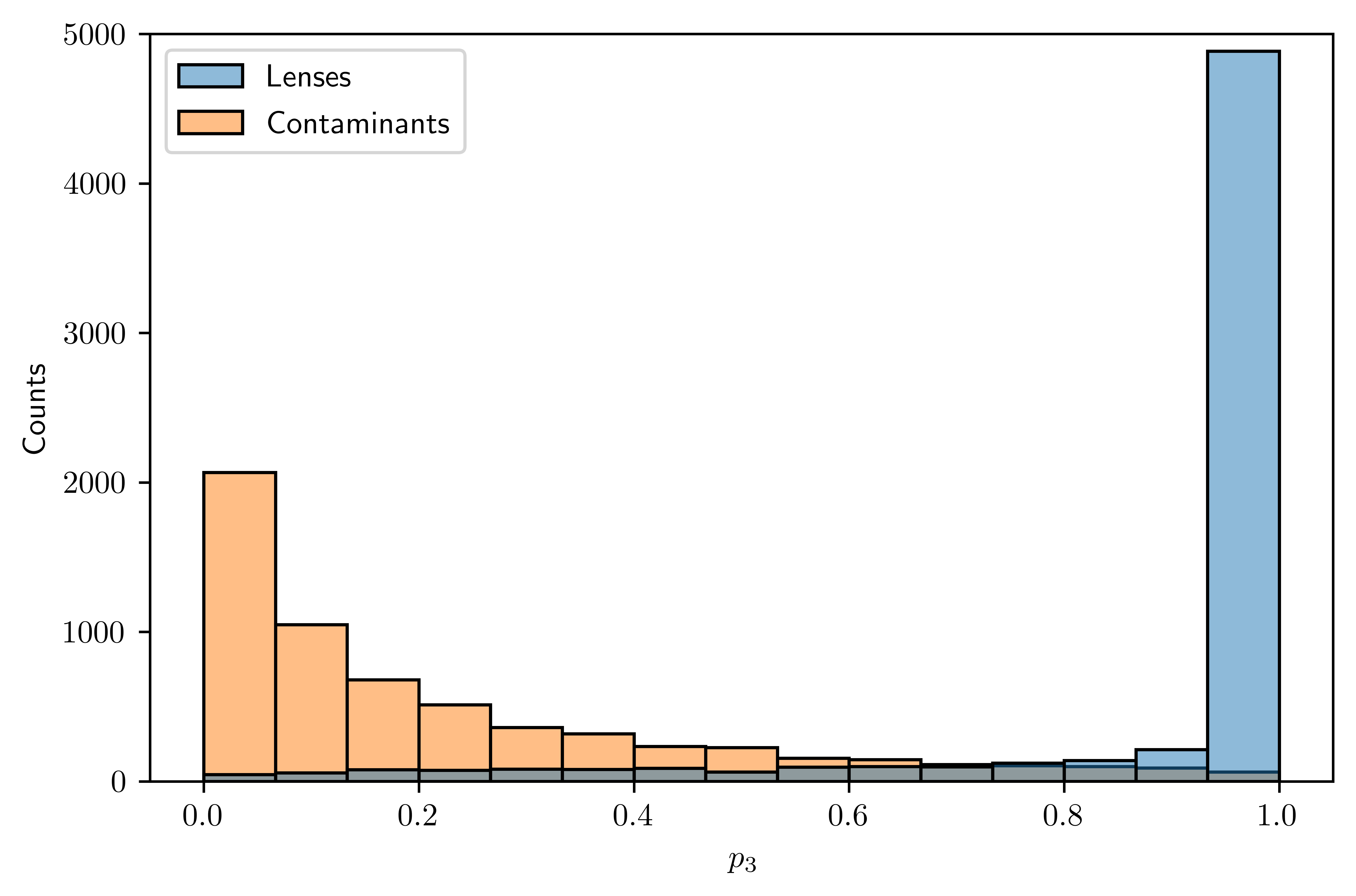}}
\caption{Distribution of \textit{single-band CNN} (a) and \textit{three-band CNN} (b) \textit{p-values} for mock gravitational lenses and contaminants in the validation set. Further details in \Sec \ref{sec:test}}
\label{fig:Hist}
\end{figure*}

\subsection{Training phase}
Once the training set has been created, we train our CNNs using the \textit{mini-batch stochastic gradient descent} technique. Each mini-batch is made up of 64 images (32 strong gravitational lenses and 32 contaminants). Our CNNs minimise the binary cross-entropy (Eq.~\ref{eq:cross_entropy}) using the \textsc{Adam} optimiser \citep{Kingma_Adam} (see \Sec\ref{sec:CNN}). We initially set \textsc{Adam}'s learning rate to $10^{-2}$, gradually lowering it up to $10^{-5}$ during the training phase to fine-tune the weights. As done in \citet{He_ResNet}, we initialise the CNNs weights $w_i$ following a normal distribution with $\mu=0$ and $\sigma=1/n$ where $n$ is the number of inputs of each unit. To increase the training set size and to teach the CNNs rotational, scaling, and translational invariance, we employ \textit{data augmentation} \citep{Simard_Augmentation}. This is a common strategy in machine learning, consisting of feeding several copies of the same image to the CNNs. Each copy is:
\begin{enumerate}
\item Rotated by an angle between 0 and 360 degrees;
\item Translated in the horizontal and vertical direction by N pixels, with N in the range [-4,4];
\item Reflected on the vertical and horizontal axis with a 50\% probability;
\item Rescaled by a factor in the range [1/1.1, 1.1].
\end{enumerate}

We directly perform data augmentation during the training phase using the opensource python library \textsc{scikit-image} \citep{Walt_Scikit}. Comparing to other analogous experiments, we employ a more limited-size training set. To prevent overfitting, we use cross-validation constantly monitoring validation loss and accuracy. We stop the training when the validation accuracy reaches its maximum at  $\sim90\%$. The CNN performs overall $\sim 40,000$ weights updates, examining $\sim 10^6$ examples in total.

\begin{figure*}
\includegraphics[scale=0.18]{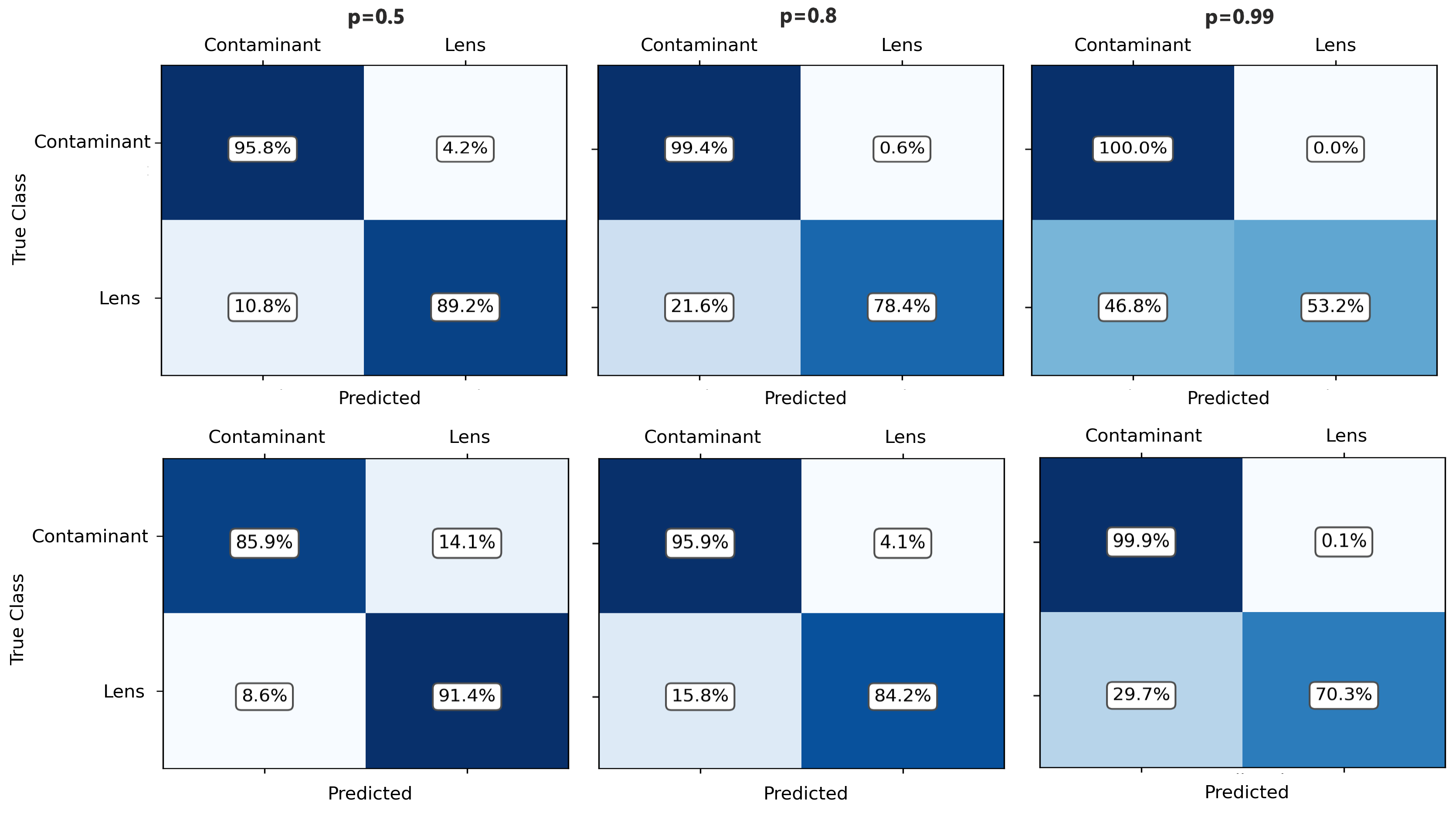}
\caption{Confusion Matrices for the \textit{single-band CNN} (top row) and \textit{three-band CNN} (bottom row). All the metrics are computed applying the lens-finders to the validation set and choosing different values of the threshold. Analysing the different matrices, we choose a threshold value of 0.8 (\Sec\ref{sec:test})}
\label{fig:conf}
\end{figure*}

\section{Testing the performance of the CNNs}
\label{sec:test}

Before applying the CNNs to real data searching for strong gravitational lenses, we need to assess their performances. We thus apply both CNNs to a validation set made up of mock gravitational lenses and contaminants. We produce mock lenses following the same procedure as discussed in \Sec \ref{sec:positive}, while we select contaminants trough the same distribution as described in \Sec \ref{sec:negative}. Each CNN assigns a \textit{p-value} between 0 and 1 to all images, related to the probability of being a strong gravitational lens. We show the \textit{p-value} distributions for mock lenses and contaminants, where the ground truth is known, in \Fig\ref{fig:Hist}. An ideal classifier would assign $p=1$ to all lenses and $p=0$ to all contaminants. We thus need statistical indices (i.e. “metrics”) to measure the difference between our lens finder and an ideal one.

\subsection{Confusion Matrix}
\label{sec:CM}
 A \textit{confusion matrix} is a table containing four values: \textit{True Positive Rate} (TPR), \textit{False Positive Rate} (FPR), \textit{True Negative Rate} (TNR), and \textit{False Negative Rate} (FNR). They are defined as follows:
 \begin{equation}
     TPR=\frac{TP}{TP+FN}
 \end{equation}
  \begin{equation}
     TNR=\frac{TN}{TN+FP}
 \end{equation}
  \begin{equation}
     FPR=\frac{FP}{TN+FP}=1-TNR
 \end{equation}
  \begin{equation}
     FNR=\frac{FN}{TN+FP}=1-TPR
 \end{equation}
 
Where FN, FP, TP and TN are, respectively, the number of False Negatives, False Positives, True Positives and True Negatives. All these values are computed once a threshold value ($p_{\rm{Th}}$) is chosen, and considering all objects with $p\ge p_{\rm{Th}}$ as valid lens candidates. An ideal classifier would have TPR$=$TNR$=1$ or, equivalently, FNR$=$FPR$=0$ for all possible threshold values that are not exactly zero or one. \Tab\ref{fig:conf} represents our CNNs confusion matrices for different values of $p_{\rm{Th}}$. As expected, for both networks the fraction of False Positives decreases towards higher $p_{\rm{Th}}$, while the fraction of False Negatives increases. Following \citet{Petrillo_2,Petrillo_3}, we choose an intermediate threshold value of 0.8 to get a fair trade-off between purity (i.e. a low number of false positives) and completeness (i.e. a low number of false negatives) for the resulting candidate sample.

\subsection{Receiver Operating Characteristic}
\label{sec:ROC}
\begin{figure}
	\includegraphics[width=\columnwidth]{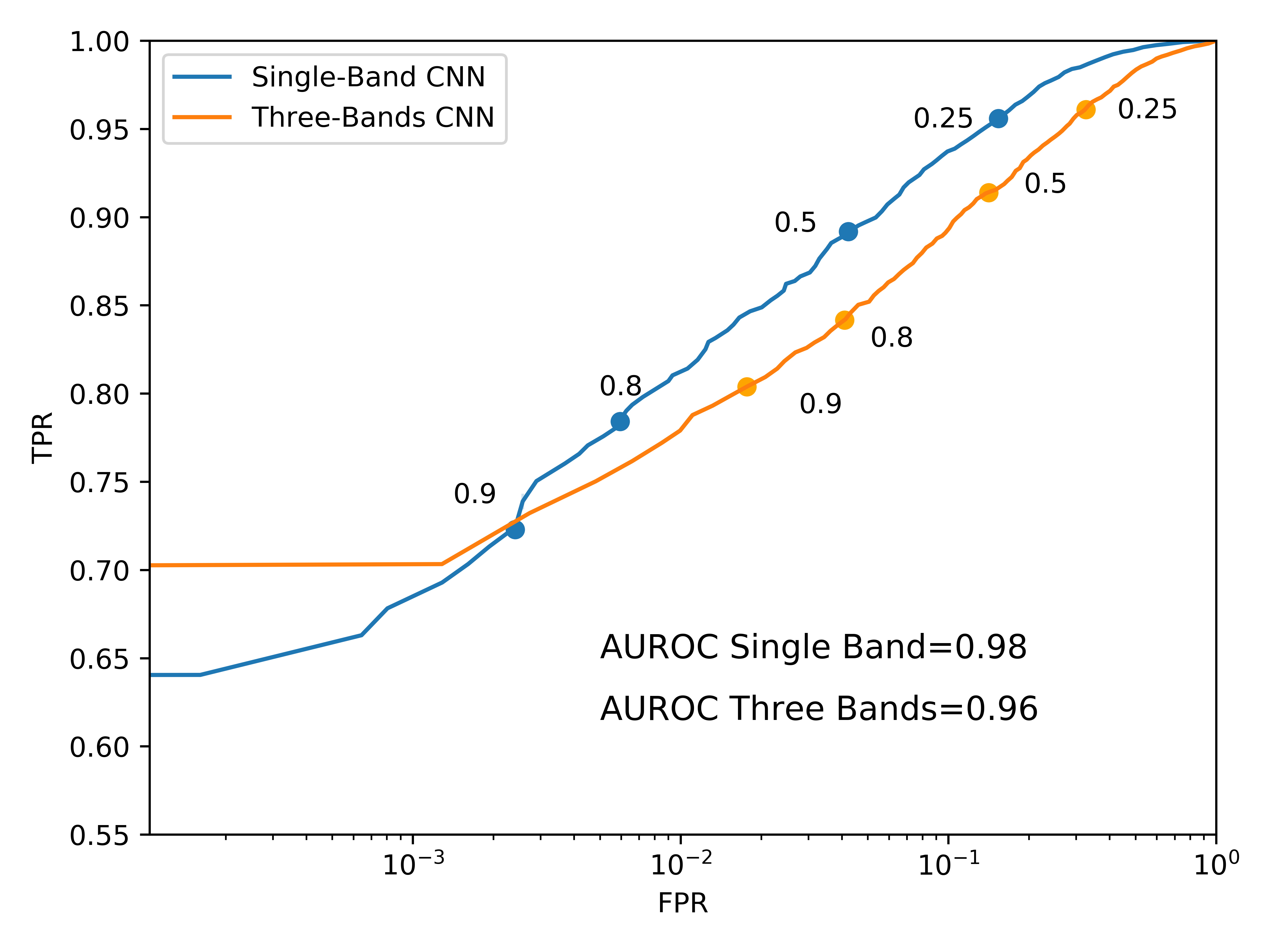}
    \caption{\textit{Receiver Operating Characteristic} curve (ROC curve) for the two CNNs built during cross-validation. The plot is in a semi-logarithmic scale to better show low \textit{False Positive Rate} (FPR) values and to show the little difference between the two curves. On the plot are reported different values of the threshold value $p_{\rm{Tresh}}$ (\Sec \ref{sec:ROC})}
    \label{fig:ROC}
\end{figure}

 A \textit{Receiver Operating Characteristic} curve (or ROC curve) is produced computing TPR and FPR for all possible threshold values and plotting them against each other. An ideal classifier would provide a ROC curve passing by the point (TPR=1, FPR=0), while a poorly-efficient one would produce a ROC curve lying on the bisector of the TPR-FPR plane. \Fig\ref{fig:ROC} represents our CNNs ROC curves. To quantitively measure the performances of a classifier, we can compute the AUROC (\textit{Area under the ROC curve}) which is equal to 1 for an ideal classifier and to 0.5 for a not-optimal one. Our \textit{single-band CNN} produces an AUROC=0.98, while the \textit{three-band CNN} produces an AUROC=0.96. Both metrics are similar to other analogous CNN-based lens finders (see, e.g., the results of the first \textit{strong lens finding challenge}; \citealt{Metcalf_Challenge}).

\subsection{$F_{\beta}$}
\label{sec:F_Beta}
We employ a third metric called $F_{\beta}$ \citep{Yates_F_beta}. This metric, commonly employed to measure performances of classification algorithms, was also used to rank the entries in the second edition of the \textit{strong lens finding challenge} (Metcalf et al., in prep). It is defined as a weighted geometric average of the precision and recall of the CNN:
\begin{equation}
    F_{\beta}=(1+\beta^2)\frac{P \times R}{\beta^2 P + R}
\end{equation}
where 
\begin{equation}
    P=\textrm{precision}=\frac{TP}{TP+FP}
\end{equation}
\begin{equation}
    R=\textrm{recall}=\frac{TP}{TP+FN}=TPR.
\end{equation}
Varying the $\beta$-factor, we can differently weight precision and recall. Since in real data non-lenses are largely more abundant than lenses, we prefer having a highly pure candidate sample rather than a highly complete one. We thus use a $\beta^2=0.001$, as in Metcalf et al. (in prep). An ideal classifier would have a maximum $F_{\beta}=1$. Our \textit{single-band CNN} reached a maximum  $F_\beta=0.9994$, while the \textit{three-band CNN} reached a maximum $F_\beta=0.9993$. As before, these values are similar to other analogous CNN-based lens finders (Metcalf et al., in prep).

\begin{figure*}
\subfloat[][]
{\includegraphics[scale=0.55]{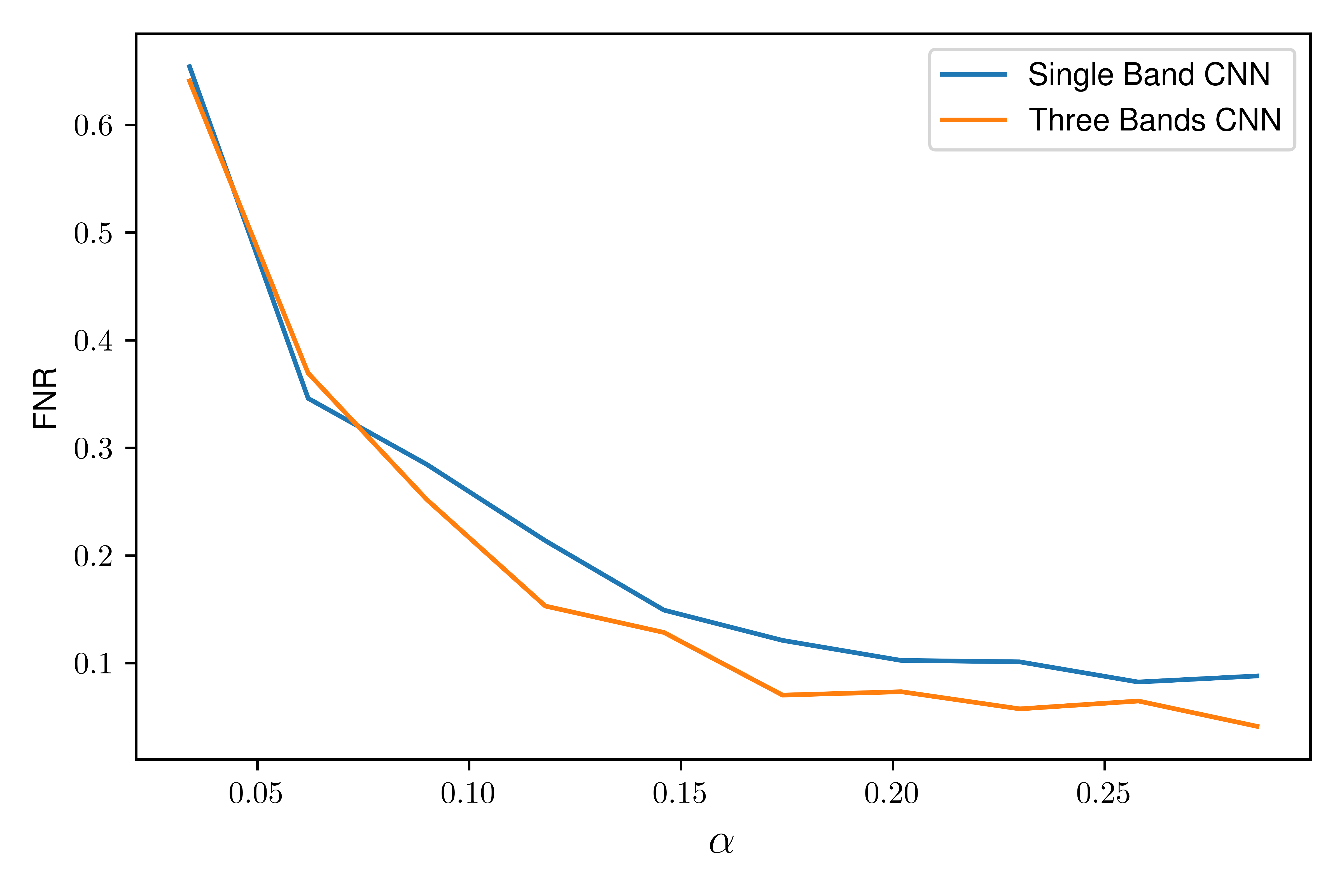}}\qquad
\subfloat[][]
{\includegraphics[scale=0.55]{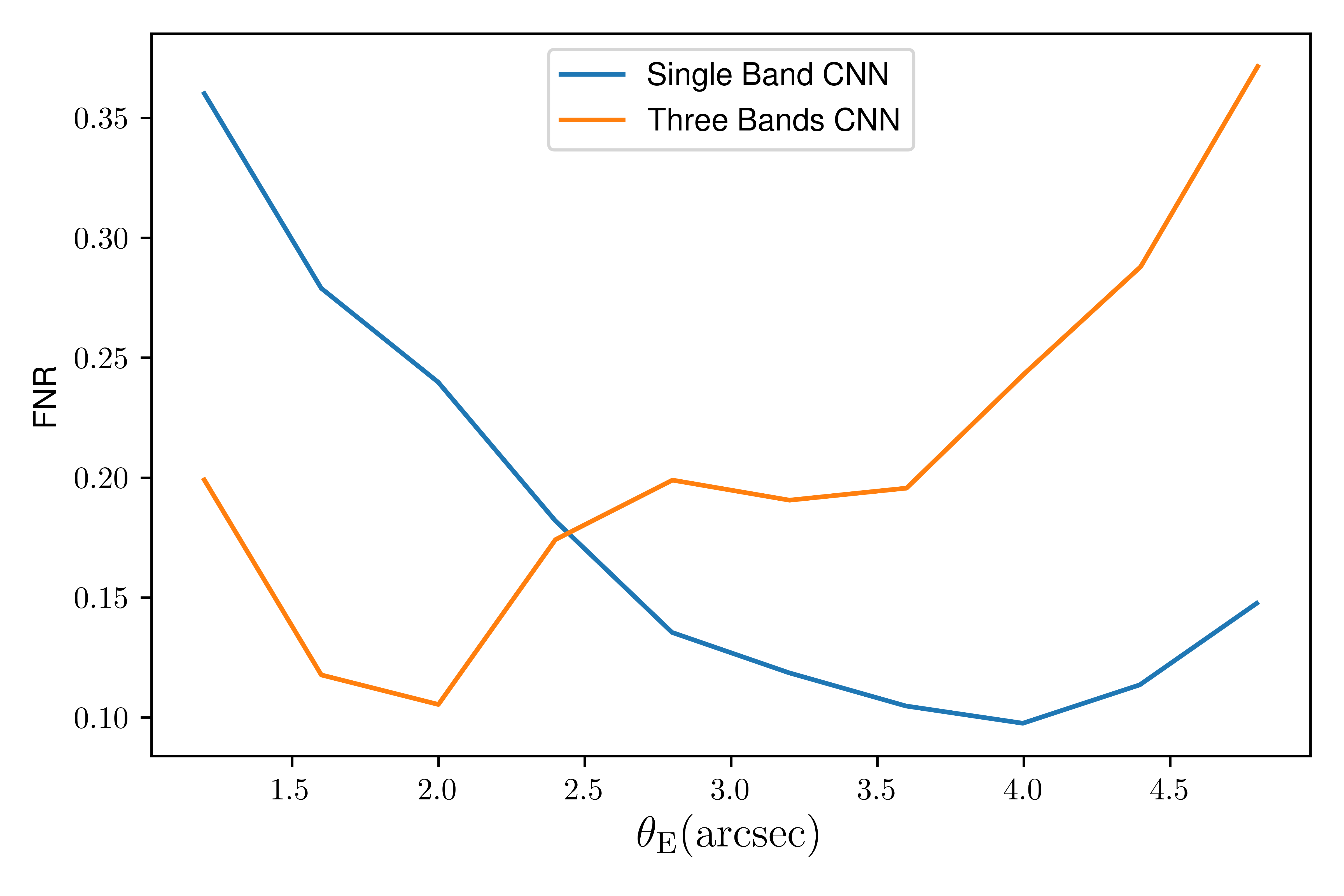}}
\caption{\textit{False Negative Rate} (FNR) as a function of the luminosity ratio between arc and deflector (a) and as a function of Einstein radius (b). FNRs are computed applying both CNNs to the validation set (\Sec \ref{sec:further}).}
\label{fig:further}
\end{figure*} 

\subsection{Further Analyses}
\label{sec:further}
It is interesting to measure performances as a function of lens parameters such as the $\alpha$-factor (that describes the brightness of the source versus the lens; see \Sec\ref{sec:positive}) or Einstein radius. \Fig\ref{fig:further} shows our results. As expected, the FNR decreases towards larger $\alpha$-factors and thus towards gravitational arcs with higher brightness. It is worth noting that the two CNNs react differently to different Einstein radii. Lenses with smaller Einstein radius often have unresolved gravitational arcs. These are harder to detect using only \textit{r-}band images. On the contrary, the \textit{three-band CNN} can more easily recognise the colour gradient between the deflector and the gravitational arc, producing a lower FNR. Conversely, lenses with larger Einstein radii more easily confuse the \textit{three-band CNN}: distant gravitational arcs are often mistaken for blue galaxies in the lens environment. \textit{Single-band CNN}, thanks to better image quality, can more easily detect the arc because of its morphology.

\subsection{Final considerations}

Analysing the different metrics, we conclude that the two CNNs are complementary (an analogous result was found by \citealt{Petrillo_2,Petrillo_3}). Although the \textit{single-band CNN} performs slightly better on the validation set analysing global metrics (ROC curve and $F_{\beta}$ index), the \textit{three-band CNN} produces a more complete candidate sample for the chosen threshold value (\Tab\ref{fig:conf}). Analysing also \Fig\ref{fig:further}(b), it can be seen that the \textit{three-band CNN} attains lower FNR for smaller Einstein radii (which are the most common, see \citealt{Collett_LensPop}). We thus decide to use both CNNs to search for strong gravitational lenses in real data produced by the VOICE survey.

\begin{figure*}
	\includegraphics[scale=0.45]{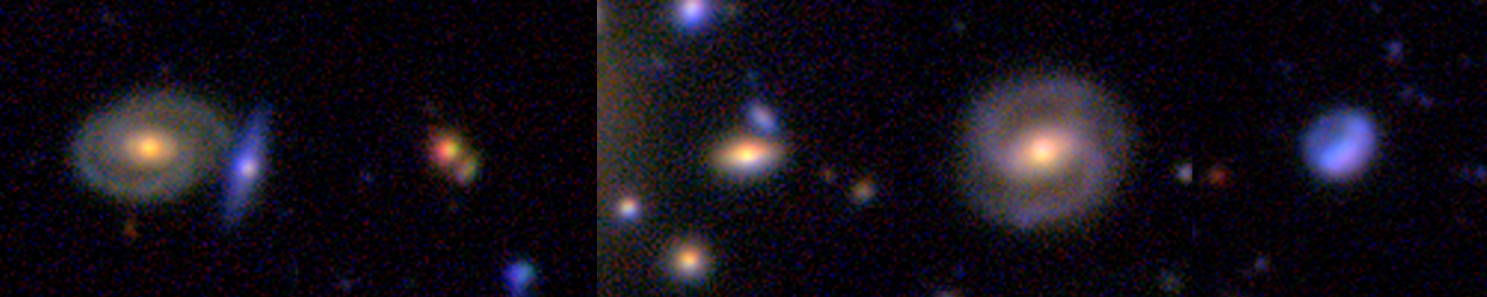}
    \caption{Some examples of contaminants wrongly classified as lenses by the CNNs and with visual score=0. Among the most common misclassified objects there are spirals, merging, galaxies with close companions and peculiar objects. All images have a 20 arcsec side. Further details in \Sec\ref{sec:Visual}}
    \label{fig:Contaminants}
\end{figure*}

\section{Results}
\label{Sec:Results}
Having assessed the performances of the two CNNs, we apply both algorithms to real data from the VOICE survey. This step has a double importance. On one hand, it allows us to assemble a sample of likely strong gravitational lenses in the Chandra Deep Field South. On the other hand, it represents a further confirmation of the ability of the CNNs to identify strong gravitational lenses in real astronomical images. Since we trained the networks only on simulated arcs, applying the CNNs to real data helps us to exclude any possible bias in the simulation procedure. 

\subsection{Application to real data}
\label{sec:Real_data}
Differently from analogous experiments \citep[e.g.,][]{Petrillo_2,Petrillo_3}, applying the CNNs to a smaller survey, we are able to search for strong lenses in a larger fraction of observed galaxies than just in the LRG sample \citep{Li_KiDS}. We analyse all the $\sim21,200$ galaxies in the Bright Galaxies Sample (see \Sec\ref{sec:Sample}) passing their 101x101 pixels$^2$ stamps in the \textit{gri} photometric bands, or only in \textit{r},  to the two CNNs. Both algorithms give as output two values ($p_1$ for the \textit{single-band CNN} and $p_3$ for the \textit{three-band CNN}) in the range $[0,1]$. Choosing a threshold value $p_{\rm{Th}}=0.8$ (\Sec\ref{sec:CM}) and considering all the images with $p>p_{\rm{Th}}$ as lens candidates, we assemble two samples: the \textit{single-band lens-candidate sample} (103 systems with $p_1>0.8$) and the \textit{three-band lens-candidate sample} (161 systems with $p_3>0.8$), which we finally join in a combined candidate sample (\textit{CNNs sample} hereafter). The full \textit{CNNs sample} consists of 257 galaxies with at least one \textit{p-value} above the chosen threshold (7 of which have both \textit{p-values} above the threshold), $\sim1\%$ of the full BG sample. 

\begin{figure}
	\includegraphics[width=\columnwidth]{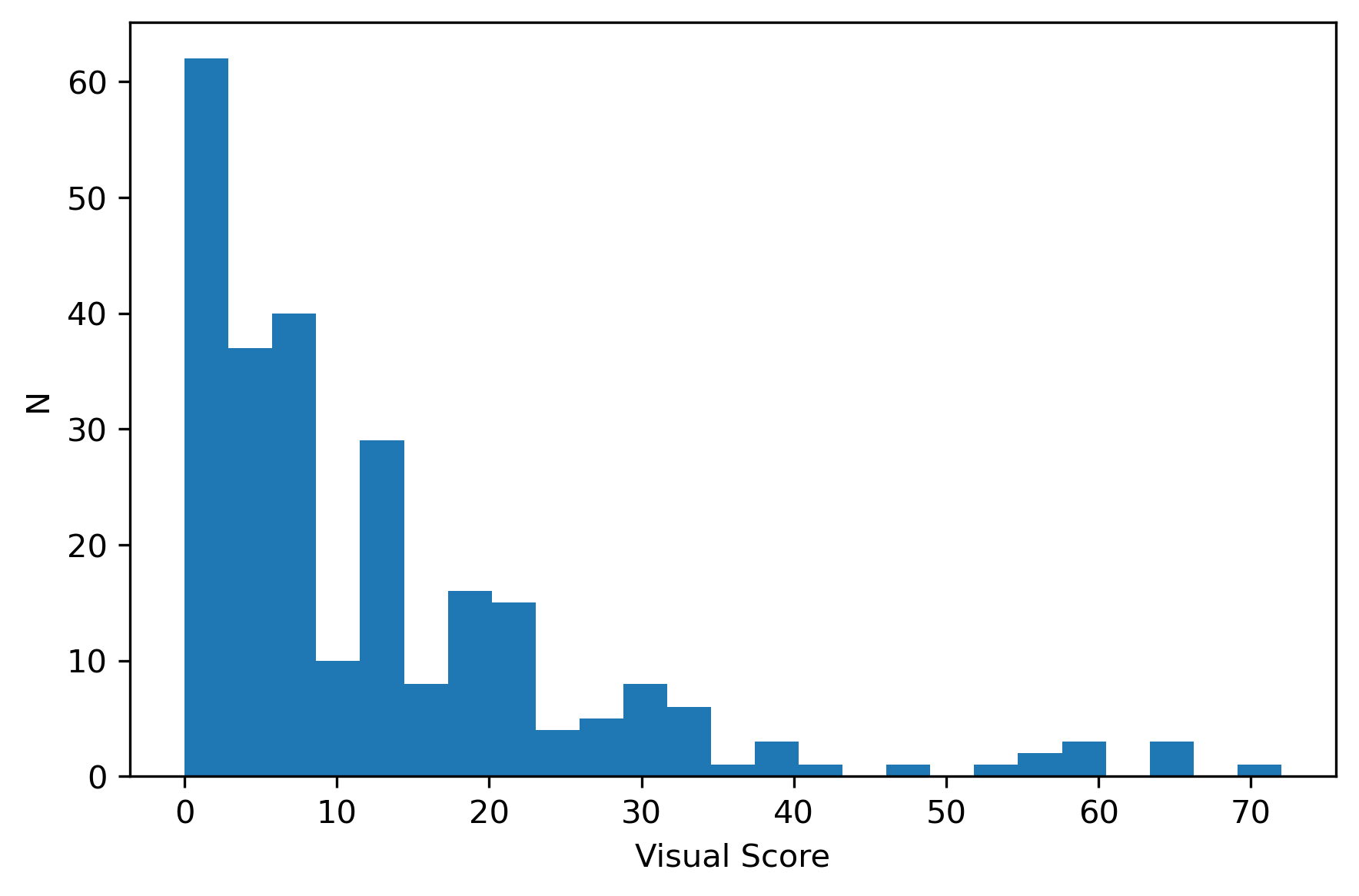}
    \caption{Results of the visual inspection performed on the 257 candidates accepted by the CNNs. We choose a threshold score of 36 to consider a system as a lens candidate. Further details in \Sec\ref{sec:Visual}}
    \label{fig:Scores}
\end{figure}

\subsection{Visual Inspection}
\label{sec:Visual}
We do not expect the CNNs to retrieve a completely pure candidate sample (see \Sec\ref{sec:test} and \Tab\ref{fig:conf}). We expect slightly lower performances passing from the validation set (made up of simulations) to real data. To further clean the final sample from false positives, we perform a visual inspection of the images retrieved by the CNNs. Nine of the authors (the \textit{graders} hereafter) inspected all the 257 images in the \textit{CNNs sample}. Each grader could choose three different quality values for each image: 

\begin{itemize}
    \item A : \textit{Sure lens}
	\item B : \textit{Maybe lens}
	\item C : \textit{Not lens}
\end{itemize}

To combine the different rankings, as in \citet{Petrillo_1,Petrillo_2,Petrillo_3}, we assign a numerical value to each grading (10 to A, 4 to B, and 0 to C). This choice allows us to weight more the \textit{sure lens} grade than the \textit{maybe lens} when combining the rankings. Although a visual inspection is still necessary to identify false positives, it is still prone to several biases. The first is the subjectivity of the visual inspection: starting from different professional experiences, each grader might have a different idea of what a lens is. Like in other studies \citep[see e.g.][]{Petrillo_1,Petrillo_2,Petrillo_3}, also in our sample there are indeed objects graded from different authors as “\textit{sure lens}” and “\textit{no lens}”. To mitigate the effects of subjectivity, we involved more than one grader and choose a threshold value smaller than 90 (i.e. nine classifications as "\textit{sure lens}"). In doing so, we include in the final sample also candidates without unanimous ranking as “\textit{sure lens}”. The second possible bias is the inter-dependence of the graders. To mitigate this effect, all graders independently rank the images using a specially-designed grading software. This also accelerates the inspection phase and avoids any accuracy loss due to a time-consuming, tedious procedure. The results of the visual inspection are summarised in \Fig\ref{fig:Scores}, where the sum of the scores of all the graders is considered for each candidate.

At the end of the visual inspection, 194 of the 257 images ($\sim$75\%) attain at least one classification as "\textit{maybe lens}". Among the most common objects with unanimous classification as "\textit{not lens}" there are spirals, merging galaxies, polar rings, and galaxies with close companions (see some examples in \Fig\ref{fig:Contaminants}). Although the number of false positives is low in our candidate sample, these objects could represent a problem for the future applications of the CNNs. Including a higher percentage of these objects in the negative training set could improve the quality of future trainings. 

\begin{figure*}
\subfloat[][LIVE-1]
{\includegraphics[scale=0.4]{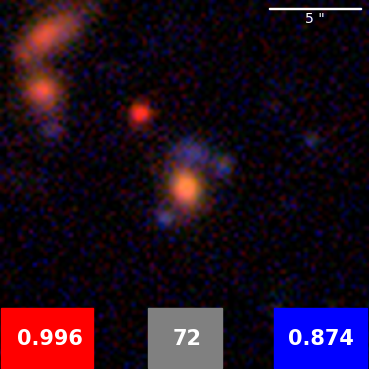}}\qquad
\subfloat[][LIVE-2]
{\includegraphics[scale=0.4]{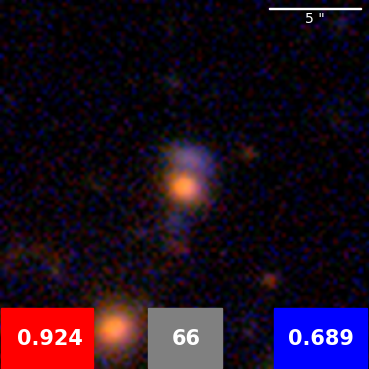}}\qquad
\subfloat[][LIVE-3]
{\includegraphics[scale=0.4]{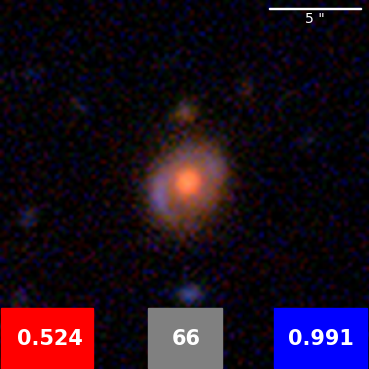}}\qquad
\subfloat[][LIVE-4]
{\includegraphics[scale=0.4]{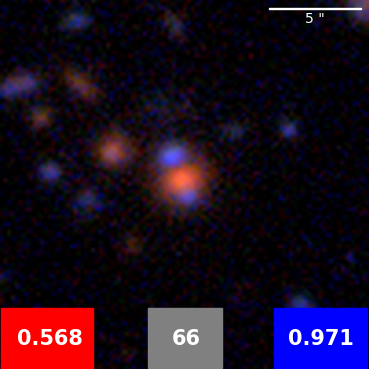}}\qquad
\subfloat[][LIVE-5]
{\includegraphics[scale=0.4]{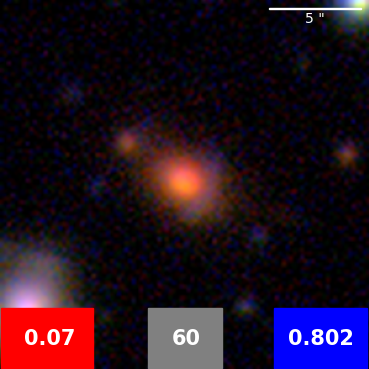}}\qquad
\subfloat[][LIVE-6]
{\includegraphics[scale=0.4]{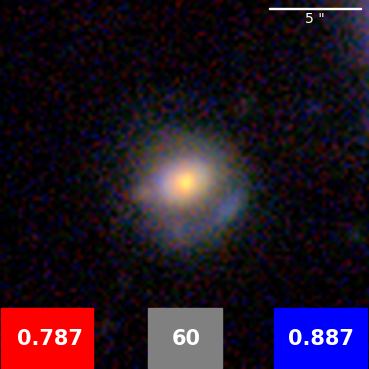}}\qquad
\subfloat[][LIVE-7]
{\includegraphics[scale=0.4]{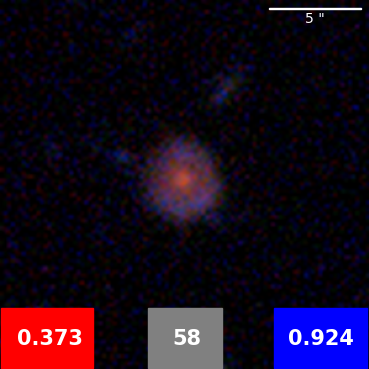}}\qquad
\subfloat[][LIVE-8]
{\includegraphics[scale=0.4]{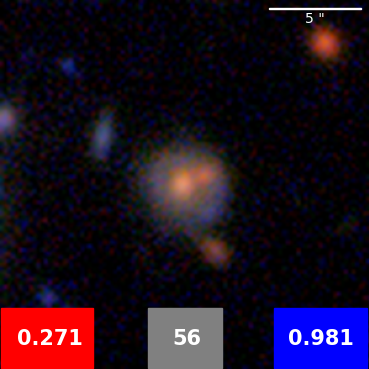}}\qquad
\subfloat[][LIVE-9]
{\includegraphics[scale=0.4]{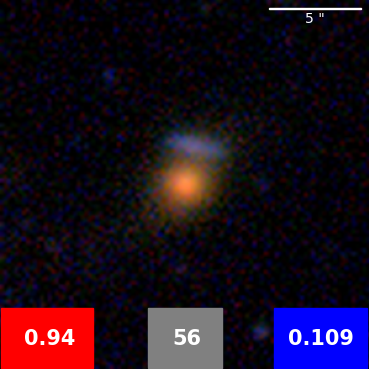}}\qquad
\subfloat[][LIVE-10]
{\includegraphics[scale=0.4]{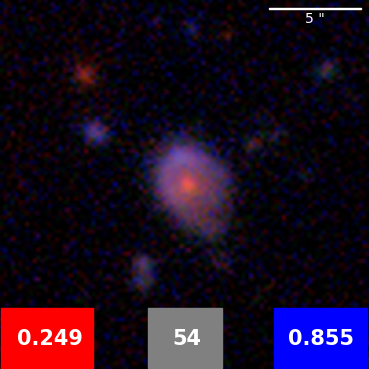}}\qquad
\subfloat[][LIVE-11]
{\includegraphics[scale=0.4]{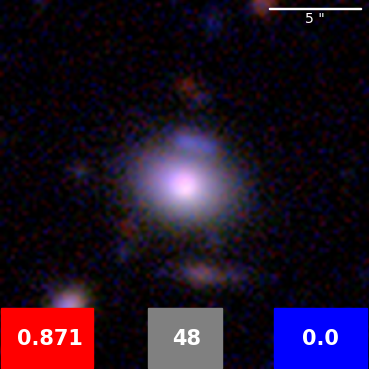}}\qquad
\subfloat[][LIVE-12]
{\includegraphics[scale=0.4]{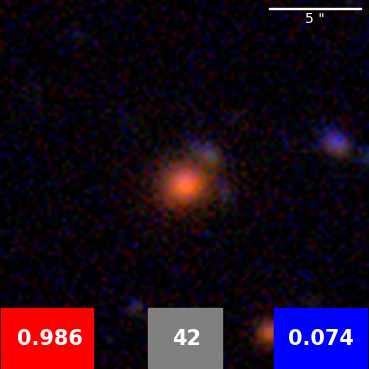}}\qquad
\subfloat[][LIVE-13]
{\includegraphics[scale=0.4]{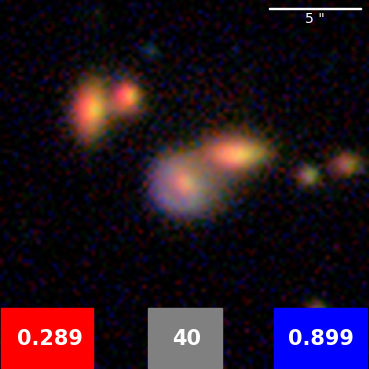}}\qquad
\subfloat[][LIVE-14]
{\includegraphics[scale=0.4]{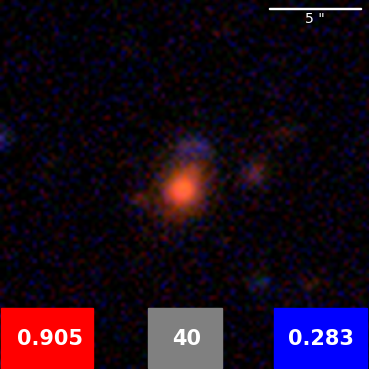}}\qquad
\subfloat[][LIVE-15]
{\includegraphics[scale=0.4]{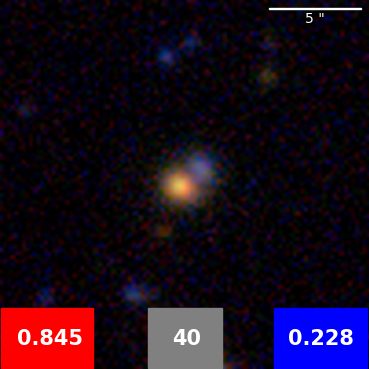}}\qquad
\subfloat[][LIVE-16]
{\includegraphics[scale=0.4]{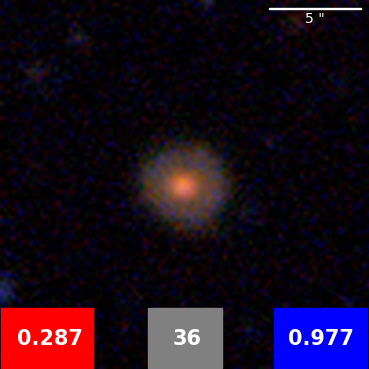}}
\caption{RGB stamps of lens candidates in the LIVE sample. Each image contains the two \textit{p-values} from the \textit{single-band CNN} (red) and \textit{three-band CNN} (blue). The systems are ordered from top to bottom according to the final score obtained from visual inspection (grey box). All stamps have a 20 arcsec side and are produced using the \textsc{HumVi} library. Further details in \Sec\ref{sec:Visual}}
\label{fig:sample}
\end{figure*}

Analysing the results, we decide to accept as lens candidates objects with a total grade $\ge 36$. This score corresponds to an unanimous classification of "\textit{maybe lens}", but includes in the final sample systems with some "\textit{not lens}" grades balanced by some "\textit{sure lens}" grades. We thus assemble the "\textit{Lenses In VoicE} (LIVE) sample, made up of 16 likely strong gravitational lenses. RGB stamps of the systems in the sample are shown in \Fig\ref{fig:sample} and listed in \Tab\ref{Tab:Candidates}.

\begin{table*}
\caption{Results of the visual inspection performed on the systems in the CNN sample (\Sec\ref{sec:Real_data}). For all the candidates with score$\ge 36$ the LIVE ID, the coordinates, the \textit{r}-band magnitude, the redshift, the \textit{p-values} from the CNNs and the visual score are provided. The last column reports if the candidate is part of the LRG sample or not (\Sec\ref{sec:Sample}). Further details in \Sec\ref{sec:Visual}}
\begin{tabular}{ccccccccccc}
\hline
LIVE ID & Name & RA & DEC & $r$ & \textit{photo-z}$^{(a)}$ & \textit{spec-z} & $p_1$ & $p_3$ & Score & LRG \\
\hline
  1 & VOICE J590934-282157 & 52.1595 & -28.3658 & 21.24 & 0.56 &  & 0.99 & 0.87 & 72 &\\
  2 & VOICE J522422-284602 & 52.4060 & -28.7672 & 20.92 & 0.56 &  & 0.92 & 0.69 & 66 &\\
  3 & VOICE J531403-273933 & 53.2342 & -27.6592 & 20.33 & 0.57 & 0.62$^{(b)}$ & 0.52 & 0.99 & 66 &\\
  4 & VOICE J534529-270652 & 53.7581 & -27.1145 & 21.47 &  &  & 0.57 & 0.97 & 66 & $\checkmark$\\
  5 & VOICE J530933-275654 & 53.1592 & -27.9482 & 20.51 & 0.61 & 0,61$^{(b)}$ & 0.07 & 0.8 & 60 &\\
  6 & VOICE J524537-275130 & 52.7603 & -27.8585 & 19.14 & 0.43 & 0.34$^{(c)}$ & 0.79 & 0.89 & 60 & $\checkmark$\\
  7 & VOICE J532836-274052 & 53.4766 & -27.6811 & 21.38 & 0.83 &  & 0.37 & 0.92 & 58 &\\
  8 & VOICE J532701-270801 & 53.4504 & -27.1337 & 20.38 &  &  & 0.27 & 0.98 & 56 &\\
  9 & VOICE J533716-275300 & 53.6210 & -27.8833 & 20.64 & 0.57 &  & 0.94 & 0.11 & 56 & $\checkmark$\\
  10 & VOICE J514601-284848 & 51.7670 & -28.8133 & 20.84 & 0.75 &  & 0.25 & 0.85 & 54 &\\
  11 & VOICE J525404-274159 & 52.9012 & -27.6998 & 18.94 & 0.14 &  0.07$^{(b)}$ & 0.87 & 0.04 & 48 &\\
  12 & VOICE J530605-280115 & 53.1013 & -28.0207 & 21.39 & 0.63 & 0.62$^{(b)}$ & 0.99 & 0.07 & 42 & $\checkmark$\\
  13 & VOICE J521812-280115 & 52.3033 & -28.0375 & 20.27 & 0.55 & 0.54$^{(d)}$ & 0.29 & 0.9 & 40 & $\checkmark$\\
  14 & VOICE J520548-282538 & 52.0968 & -28.4273 & 21.38 & 0.64 &  & 0.91 & 0.28 & 40 & $\checkmark$\\
  15 & VOICE J515150-272926 & 51.8638 & -27.4905 & 21.09 & 0.55 &  & 0.85 & 0.23 & 40 &\\
  16 & VOICE J534638-271022 & 53.7771 & -27.1727 & 20.76 &  &  & 0.29 & 0.98 & 36 & \\
\hline
\end{tabular}
\begin{flushleft}
{\footnotesize $^{\mathrm (a)}$ photometric redshift estimated with \textsc{Metaphor} \citep{cavuoti_metaphor} \\ $^{\mathrm (b)}$ spectroscopic redsift retrieved from the ACES survey \citep{cooper_aces} \\ $^{\mathrm (c)}$ spectroscopic redshift retrieved from \citet{Cowie_GALEX} \\$^{\mathrm (d)}$ spectroscopic redshift retrieved from the BLAST survey \citep{Eales_BLAST}}
\end{flushleft}
\label{Tab:Candidates}
\end{table*}

\begin{figure}
    \centering
	\includegraphics[scale=0.5]{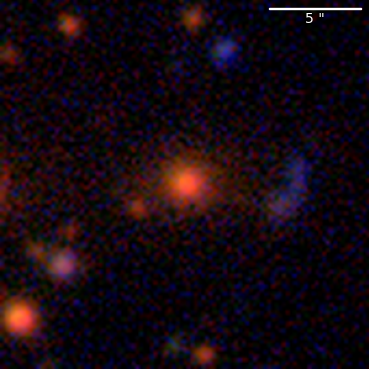}
    \caption{RGB stamp of the gravitational lens DES J0329-2820 previously discovered by \citet{Nord_DESLens} observed in the VOICE survey. The system has \textit{r}>21.5, thus it is not part of the BG sample. We manually pass its image to the CNNs, receiving both \textit{p-values} below the chosen threshold. The high value of the Einstein radius ($7.8''\pm1.4''$) can explain this result. Further details in \Sec\ref{Sec:Discussion}}
    \label{fig:lens}
\end{figure}

\section{The LIVE Sample}
\label{Sec:Discussion}
In this section, we analyse the gravitational lens candidates in the LIVE sample, assembled in the previous section. Among those systems, seven candidates were identified by the \textit{single-band CNN} and ten by the \textit{three-band CNN} (\Tab\ref{Tab:Candidates}). Only one object (LIVE-1, attaining the highest score in the visual inspection) passed the \textit{p-value} threshold for both CNNs. This result confirms the expected performances of the CNNs discussed in \Sec\ref{sec:test}. The \textit{three-band CNN} was indeed expected to retrieve a more complete candidate sample (i.e., to identify a larger fraction of real lenses), while the \textit{single-band CNN} was expected to retrieve a purer one (i.e., with a smaller fraction of contaminants). However, we highlight that, in the final candidate sample, there are objects with one \textit{p-value} nearly close to zero (LIVE-5 for the \textit{single-band CNN} and LIVE-11 for the \textit{three-band CNN}). This represents a further confirmation of the complementarity of the two algorithms (\Sec\ref{sec:test}). 

\begin{figure*}
	\includegraphics[scale=0.5]{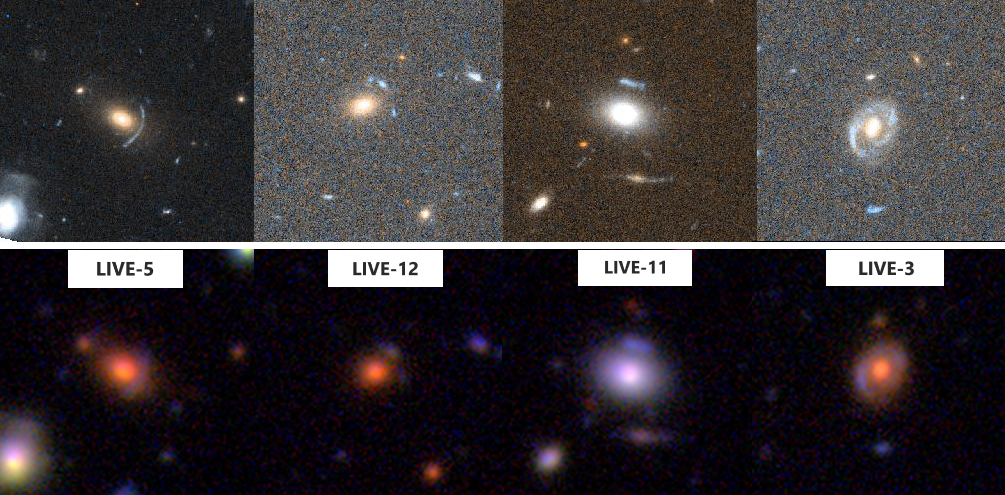}
    \caption{Systems LIVE-5, LIVE-12, LIVE-11 and LIVE-3 observed by the Hubble Space Telescope (top row) and in the VOICE survey (bottom row). The HST images are part of the GEMS survey (HST observing program 9500, \citealt{rix_gems}) and are retrieved from the Hubble Legacy Archive. All the images have a 20 arcsec side. LIVE-5 is a previously discovered strong gravitational lens \citep{blakeslee_lens}. LIVE-12 and LIVE-11 show likely lensing features, while LIVE-3 shows a likely spiral structure. Further details in \Sec\ref{Sec:Discussion}.}
    \label{fig:HST_1}
\end{figure*}

Among the systems in the LIVE sample, there is a previously discovered gravitational lens (LIVE-5, \citealt{blakeslee_lens}). It is worth noting that this object did not attain a unanimous classification as “\textit{sure lens}” during the visual inspection, albeit it did not obtain any "\textit{not lens}” classification. This confirms the possible biases in the visual inspection (\Sec\ref{sec:Visual}) and exemplifies the dependence of the classification on the image quality and the signal-to-noise ratio (\Fig\ref{fig:HST_1}). Furthermore, it is interesting to note that this real lens has $p_1$ below the chosen threshold. This can be explained by the higher FNR of the \textit{single-band CNN} at lower values of the Einstein radius (\Sec\ref{sec:test} and \Fig\ref{fig:further}). Through research in the current literature, we retrieve one more lens previously identified in the CDFS (DES J0329-2820; \citealt{Nord_DESLens}). However, since this system has a \textit{r}-band magnitude of 22.4, it is too faint to be part of the BG sample (\Sec\ref{sec:Sample}), and thus it is not analysed by the CNNs. Nevertheless, we manually pass its image to the CNNs, obtaining both \textit{p-values} below the chosen threshold. This result can be explained by the large value of the Einstein estimated by \citet{Nord_DESLens} for this system ($\theta_E=7.8''\pm1.4''$, see \Fig\ref{fig:lens}). This value is well outside the range of Einstein radii on which we trained our algorithms and in which we expect the CNNs to be accurate.

Furthermore, for four of the objects in the LIVE sample, we retrieve high-resolution imaging from the Hubble Space Telescope Legacy Archive (HLA\footnote{https://hla.stsci.edu/}). One of the objects is LIVE-5 (previously discovered, \Fig\ref{fig:HST_1}). Other two objects, LIVE-11 and LIVE-12, show likely lensing features when observed with HST. On the contrary, high resolution data for LIVE-3 makes it possible to identify a likely spiral structure, revealing a non-lens nature for this candidate.

Continuing the analysis, we emphasise  that six of the candidates in the LIVE sample are part of the LRG sample (see. \Tab\ref{Tab:Candidates} and \Sec\ref{sec:Sample}), while none of the systems satisfies the criteria exposed by \citet{Eisenstein_LRG} to select LRGs. Using these criteria to select the input sample for the CNNs (as done in analogous studies, e.g. \citealt{Petrillo_1,Petrillo_2,Petrillo_3}) we would therefore have missed all these candidates.
 
Finally, to fully characterise our set of candidates, we retrieve spectroscopic and photometric redshifts for most of the systems in the LIVE sample. These values are summarised in \Tab\ref{Tab:Candidates}. The photometric redshifts are computed using the \textsc{Metaphor} algorithm \citep{cavuoti_metaphor}, previously applied to the galaxies in the VOICE survey. The spectroscopic redshifts are retrieved from the VizieR archive\footnote{https://vizier.u-strasbg.fr/} \citep{ochsenbein_vizier} querying the catalogues from previous spectroscopic surveys of the CDFS \citep{cooper_aces,Eales_BLAST,cowie_redshift}.

\subsection{Comparison with LensPop}
\label{sec:lenspop}

We employ the lens-statistics code \textsc{LensPop} to assess the reliability of the LIVE sample. \textsc{LensPop} is a software introduced by \citet{Collett_LensPop} and able to simulate a realistic population of strong gravitational lenses. By opportunely tuning its parameters, \textsc{LensPop} can predict the number of lenses observable in a given survey, and their global properties. \textsc{LensPop} defines as “observable” all the lenses satisfying these criteria:
\begin{equation}
\label{eq:lenspop}
\begin{cases}
    \theta_E^2 \ge x_s^2 + y_s^2 \\
    \theta_E^2 \ge r_s^2 + (s/2)^2 \\
    \mu_{\rm{TOT}}r_s>s , \quad \mu_{\rm{TOT}}>3 \\
    SNR \ge 20
\end{cases}
\end{equation}
Where $x_s$, $y_s$ and $r_s$ represent, respectively, the coordinates and the size of the unlensed source. $\theta_E$ represents the Einstein radius, $\mu$ the magnification, $SNR$ the signal-to-noise ratio and $s$ the mean seeing of the image \citep{Collett_LensPop}.
Besides these properties, we require that
\begin{equation}
\begin{cases}
    mag_r<21.5 \\
    1''<\theta_E<5''
\end{cases}
\end{equation}
The first property requires the galaxy being in the BG sample (and, thus, being analysed by the CNNs, \Sec\ref{sec:Sample}). The second property considers that, since we trained our algorithms on lenses with Einstein radii between 1 and 5 arcsec (\Tab\ref{tab:param}), we do not expect our CNN to be accurate outside this range \citep{Petrillo_2,Petrillo_3}.
\textsc{LensPop} predicts that 10 strong lenses are observable in the 4.9 deg$^2$ of the CDFS covered by the VOICE survey. Assuming a Poissonian noise on the code prediction, we estimate a confidence interval of $10\pm 3$. Comparing this value with the size of the LIVE sample, we expect the latter to be nearly complete, but not entirely pure. This result agrees with the likely spiral galaxy identified in the sample using the HST data (\Sec\ref{Sec:Discussion}). The lens population simulated by \textsc{LensPop} is predicted to have a mean redshift of 0.4 with a standard deviation of 0.2. Using the spectroscopic and photometric redshifts retrieved in the previous section for the LIVE sample, we estimate a mean redshift of 0.5 with a standard deviation of 0.2.
Finally, using \textsc{LensPop}, we predict a mean value of the Einstein radii for the lenses observable in the VOICE survey equals to 1.4'' with a standard deviation of 0.3''. Visually estimating the Einstein radii for the lenses in the LIVE sample as half the distance between the alleged multiple lensed images\footnote{This estimate is not completely accurate for strongly asymmetric lenses, a complete modelling would be required for a better accuracy}, we obtain a mean value of 1.5'' with a standard deviation of 0.4''.
Both the redshifts and the Einstein radii are consistent, within $1\sigma$ error, with the predictions made by \textsc{LensPop}, representing a further confirmation of the reliability of the candidates in the LIVE sample.

\subsection{Comparison with KiDS}
\label{sec:kids}

\begin{figure*}
	\includegraphics[scale=0.3]{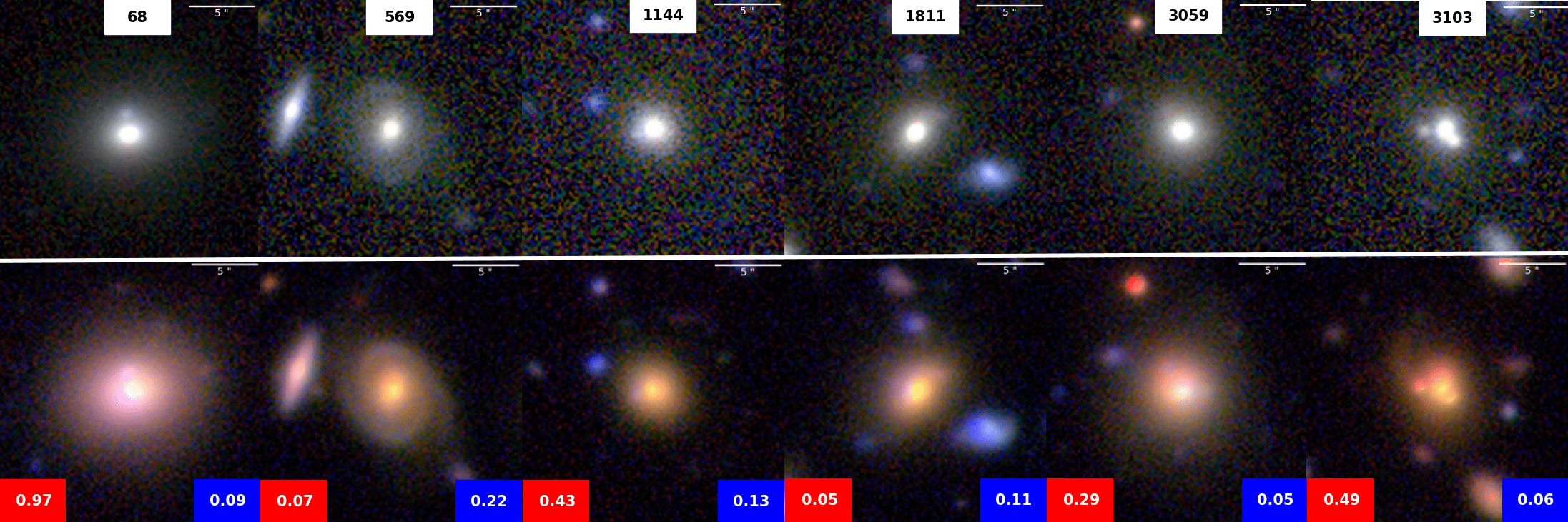}
    \caption{Six lens candidates found in the KiDS survey (top row) by \citet{Petrillo_3} and observed in the VOICE survey (bottom row). The higher SNR of VOICE reveals the non-lensing nature of all these candidates and makes it easier for the CNNs to reject these systems. For all the systems we show the LinKS ID (white box) and the two \textit{p-values} obtained by the \textit{single-band CNN} (red box) and the \textit{three-band CNN} (blue box). All the stamps have a 20 arcsec side. Further details in \Sec\ref{sec:kids}}
    \label{fig:KiDS}
\end{figure*}

The CNNs employed in this work were previously applied to data from the \textit{Kilo-Degree Survey} (KiDS, \citealt{Kuijken_KiDS,Kuijken_DR4}) by \citet{Petrillo_1,Petrillo_2,Petrillo_3}. It is interesting to compare those results with ours, to investigate what performances the same CNN architectures can achieve when applied to different data, although both produced by the same telescope, and thus similar. The images passed to the CNNs for VOICE and KiDS have the same pixel size (0.2 arcsec/pixel) and a comparable mean value of the PSF FWHM (0.8'' for VOICE \textit{r}-band, and 0.7'' for the same band in KiDS; \Sec\ref{sec:Data}, \citealt{Kuijken_DR4}). However, KiDS is a wide and shallow survey (about 900 deg$^2$ observed in the fourth data release with a 5$\sigma$ limiting magnitude in the \textit{r}-band of 25.0; \citealt{Kuijken_DR4}). VOICE, on the contrary, is a smaller but deeper survey (4.9 deg$^2$ observed with a 5$\sigma$ limiting magnitude in the same band of 26.1, see \Sec\ref{sec:Data}). 
In \citet{Petrillo_3}, the CNNs analysed 88,327 LRGs selected using the criteria exposed by \citet{Eisenstein_LRG}. The LRG sample covered less than 0.01\% of the full KiDS catalogue. Adopting the same threshold used in this paper ($p_{\rm{Th}}$=0.8), the CNNs retrieved a sample of 3500 systems (about 4\% of the LRG sample) with at least one \textit{p-value} above the threshold. Finally, performing a visual inspection of the selected candidates similar to the one exposed in this paper (with seven graders and three possible grades), the authors assembled the LinKS (Lenses In KiDS) sample\footnote{https://www.astro.rug.nl/lensesinkids/} composed by 1,983 likely strong gravitational lenses with at least one \textit{p-value} above the threshold ad at least one classification as “\textit{maybe lens}” or “\textit{sure lens}”. 89 of those candidates (the “\textit{bona fide}” sample) attained a visual score $\ge28$\footnote{\citet{Petrillo_3} used the same numerical values given in \Sec\ref{sec:Visual}}, equivalent to our threshold of unanimous classification as “maybe lens” (considering seven inspectors instead of our nine, see \Sec\ref{sec:Visual}).
In comparing the results from the two studies, we must consider two main differences between the surveys: the higher number of galaxies observed in KiDS and the higher SNR (and fainter limiting magnitude) of VOICE. Since we worked on a smaller survey\footnote{\citet{Petrillo_3} analysed $\sim900$ deg$^2$, while we only mapped 4.9 deg$^2$}, we could relax the criteria of \citet{Eisenstein_LRG} for the selection of the galaxies to analyse. The BG sample inspected by the CNNs covered a larger fraction of the galaxies observed in the VOICE survey (about 3\% against the 0.01\% of KiDS). However, the fractions of systems retrieved by the CNNs are quite similar (1\% of the BG sample in VOICE, 3\% of the LRG sample in KiDS). Performing the visual inspection, we could independently assess the contamination rate of the candidate samples. In \citet{Petrillo_3}, $\sim$57\% of the candidates attained at least one classification as “\textit{maybe lens}” or “\textit{sure lens}”. This fraction is higher in our case ($\sim$75\%, see \Sec\ref{sec:Visual}). This result can be explained by the fainter limiting magnitude reached by the VOICE survey. This allows the CNNs to identify more easily faint characteristics (e.g., spiral structures), revealing the contaminant nature of some lens candidates. This point can be studied in detail, since the KiDS and VOICE fields overlap in a region of $\sim2$ deg$^2$. In particular, six lens candidates in the LinKS sample (but with visual score $<28$, i.e. not part of the "bona fide sample") are also in the BG sample analysed by our CNNs. KiDS and VOICE cutouts for these systems are shown in \Fig\ref{fig:KiDS}. The higher SNR and better image quality of VOICE reveal the contaminant nature of all these systems. Five of these objects obtained both \textit{p-values} under the chosen threshold from our CNNs. Only the object with LinKS $\rm ID = 68$ (the second system in \Fig\ref{fig:KiDS}) obtained $p_1=0.9$ from our \textit{single-band CNN}. This candidate, however, attained a visual score of 22 during our visual inspection, well below the chosen threshold of 36.

The fainter limiting magnitude (and the consequent higher SNR), combined with more flexible criteria to select the galaxies to analyse, is also responsible for the higher number density of lenses found by our CNNs. In fact, according to equation \ref{eq:lenspop}, the higher SNR reached by a deep survey augments the number of strong lenses retrievable in a given area \citep{Collett_LensPop}. To assess quantitatively this property, we can consider the $\sim 2$ deg$^2$ area observed in both surveys. While the "bona fide" LinKS sample contains no lens candidates in this area, the LIVE sample contains six candidates (LIVE IDs: 1, 5, 6, 12, 13 and 14) retrieved by our CNNs in this region. 

Finally, a further interesting comparison between KiDS and VOICE concerns the mean redshift of the retrieved lens candidates. The systems in the LinKS sample have a mean redshift of 0.3 \citep{Petrillo_3}, while the systems in the LIVE sample have a mean redshift of 0.5 (\Sec\ref{Sec:Discussion}). Identifying strong lenses at higher redshift is crucial for extending many analyses on a larger scale \citep[see e.g.,][]{treu_DM,Treu_IMF,Koopmans_DM,Vegetti_DM}. All these results justify the increasing interest in the forthcoming deep surveys conducted with the \textit{Euclid} satellite \citep{Laureijs_Euclid} and the \textit{Vera Rubin Observatory} \citep{LSST}.

\section{Conclusion}
\label{sec:conclusion}
In this paper, we presented a sample of 16 likely strong gravitational lenses identified in the Chandra Deep Field South (CDFS). We analysed the data from the \textit{VST Optical Imaging of the CDFS and ES1 Fields} (VOICE survey, \citealt{Vaccari_VOICE}) using two Convolutional Neural Networks (CNNs). 

Both algorithms were previously developed by \citet{Petrillo_1,Petrillo_2} and employed to search for strong lenses in the \textit{Kilo-Degree Survey} (KiDS, \citealt{Kuijken_KiDS,Kuijken_DR4}) by \citet{Petrillo_1,Petrillo_2,Petrillo_3} and the \textit{Fornax Deep Survey} (FDS, \citealt{Iodice_FDS}) by \citet{Cantiello_FDS}. We trained the CNNs on composite images obtained by superimposing simulated gravitational arcs on real LRGs observed in VOICE  (\Sec\ref{sec:positive}). The first CNN, \textit{single-band CNN}, analysed images in the \textit{r} photometric band, while the second one, \textit{three-band CNN}, inspected composite RGB images obtained combining the data in the \textit{gri} bands with the \textsc{HumVi} library. Once the algorithms have been trained, we assessed their performances applying them to a validation set consisting of both simulated lenses and real contaminants (\Sec\ref{sec:test}). The performances of both networks (i.e., the False Positive Rate and the True Negative Rate) are comparable to the previous applications in \citet{Petrillo_2,Petrillo_3}. Moreover, we found that the \textit{three-band CNN} can identify more easily systems with smaller Einstein radii, where the colour gradient can help to recognise unresolved gravitational arcs. On the contrary, the \textit{single-band CNN} shows a better accuracy in identifying systems with larger Einstein radii. In this case, however, high-\textit{z} groups of star forming galaxies can be more easily mistaken for distant gravitational arcs .

Concluding that the two CNNs are complementary, we applied both networks to real data from the VOICE survey. The CNNs analysed in total $\sim 21,200$ galaxies with $mag_r<21.5$, retrieving a sample of 257 lens candidates with at least one \textit{p-value} above the chosen threshold of 0.8 (\Sec\ref{sec:Real_data}). To improve the purity of the candidate sample, we performed a visual inspection with nine graders judging the systems in a blind way (\Sec\ref{sec:Visual}). About 75\% of the candidates attained at least one classification as "\textit{maybe lens}" or "\textit{sure lens}". 

Finally, we assembled the “LIVE sample” (Lenses In VoicE) consisting of 16 likely strong gravitational lenses with at least one \textit{p-value} above the threshold and a visual score $\ge$36. To fully characterise the final set, we retrieved spectroscopic and photometric redshifts for most of the lens candidates. We also retrieved high-resolution data from the Hubble Legacy Archive for four of the systems (\Sec\ref{Sec:Discussion}). The entire process described here allowed us to identify a gravitational lens previously discovered in the CDFS \citep{blakeslee_lens} and at least two very high-probability candidates when observed by HST (\Fig\ref{fig:HST_1}). To assess the reliability of the LIVE sample, we compared its global properties with the ones predicted by the lens-statistics software \textsc{LensPop} \citep{Collett_LensPop}. We concluded that our sample is likely to be complete albeit not totally pure, while its global properties fully encompass the code predictions (\Sec\ref{sec:lenspop}). Finally, we compared our results with the ones presented in \citet{Petrillo_3}, obtained using the same CNNs applied to the KiDS survey. Since we applied the algorithms to a smaller but deeper survey, we were able to retrieve a less contaminated candidate sample, with a higher number density of lens candidates and a higher mean redshift (\Sec\ref{sec:kids}).

Although the probability to be confirmed as lens is high for most of the objects in the LIVE sample, we stress that an unambiguous validation requires a high-resolution and/or a spectroscopic follow-up \citep[see e.g.,][]{Bolton_SLACSFirst,anguita_STRIDES1,lemon_STRIDES2,spiniello_Spectrum1,spiniello_Spectrum2}, which will be provided by the \textit{Vera Rubin Observatory} deep survey that will observe the CDFS in the near future \citep{LSST}.

In conclusion, this work represents a further confirmation of the ability of machine learning algorithms like CNNs to analyse efficiently large amounts of data searching for strong gravitational lenses. These algorithms will reach their full scientific potential in the analysis of forthcoming large sky surveys such as the one performed with the ESA’s Euclid satellite \citep{Laureijs_Euclid}, the Vera Rubin Observatory \citep{LSST}, and the Chinese Space Station \citep{Gong_CSSOC}. These surveys are indeed expected to retrieve $\sim 10^5$ strong gravitational lenses in a dataset of $\sim10^9$ observed galaxies. Solely visually inspecting all the galaxies retrieved by these forthcoming facilities would require several years and would be prone to several biases (\Sec\ref{sec:Visual}), even applying some \textit{a-priori} cut to select only galaxies with a high lensing cross-section. However, even applying CNNs to select the most promising lens candidates, a low contamination rate is still crucial to reduce the need for a visual inspection. This goal can be achieved, on one hand, by employing the latest CNN architectures available \citep[see e.g.,][]{Szegedy_Inception,Chollet_Xception}, and thus taking advantage from the latest results in machine learning and computer vision. On the other hand, training these algorithms requires reliable strong lensing simulations to avoid possible biases in the training phase. This is the reason why, in the last few years, some collaborations started to investigate possible alternatives to the supervised-learning paradigm. Unsupervised learning (requiring no training set or a small one just for labelling \citep[see e.g.][]{Cheng_unsupervised} or self-supervised learning (requiring smaller datasets, e.g. \citealt{Hayat_selfsupervised}) can allow training based only on real observed strong lenses.

Finally, it is worth noting that the large amount of lenses retrieved from these forthcoming large surveys will pose the non-trivial problem of how efficiently one can then analyse and model these systems to constrain structural parameters of the lens to be used for scientific purposes. Classic bayesian techniques \citep[e.g.][]{Jullo_Lenstool,Birrer_lenstronomy,nightingale_Autolens} are indeed poorly efficient when applied to large datasets because of their computational cost and the need for human intervention. Machine learning algorithms like CNNs have already been applied to the fast automated analysis of strong gravitational lenses \citep{Hezaveh_modelingCNN,pearson_ModellingCNN,Schuldt_ModellingCNN,Madireddy_ModellingCNN}. This represents a future perspective of this work, toward a full exploitation of the scientific potential of forthcoming facilities like \textit{Euclid}, the \textit{Vera Rubin Observatory} and the \textit{Chinese Space Station}

\section*{Acknowledgements}
CS is supported by a Hintze Fellowship at the Oxford Centre for Astrophysical Surveys, which is funded through generous support from the Hintze Family Charitable Foundation. \\
MV acknowledges support from the Italian Ministry of Foreign Affairs and
International Cooperation (MAECI Grant Number ZA18GR02) and the South African
Department of Science and Innovation's National Research Foundation (DSI-NRF
Grant Number 113121) as part of the ISARP RADIOSKY2020 Joint Research Scheme.\\
LPF acknowledges the support from NSFC grants 11933002, STCSM grant 18590780100, and the Dawn Program 19SG41 \&  the Innovation Program 2019-01-07-00-02-E00032 of SMEC. \\
Based on observations made with the NASA/ESA Hubble Space Telescope, and obtained from the Hubble Legacy Archive, which is a collaboration between the Space Telescope Science Institute (STScI/NASA), the Space Telescope European Coordinating Facility (ST-ECF/ESA) and the Canadian Astronomy Data Centre (CADC/NRC/CSA).\\
This research has made use of the VizieR catalogue access tool, CDS, Strasbourg, France (DOI: 10.26093/cds/vizier).

\section*{Data Availability}
The data that support the findings of this study are available from the corresponding author, GC, upon reasonable request.

\bibliographystyle{mnras}
\bibliography{SL_Voice}

\bsp	
\label{lastpage}
\end{document}